\documentclass[twocolumns,traditabstract,longauth]{aa}

\usepackage{changepage}
\usepackage{import}
\usepackage{amsmath}
\usepackage{tabularx,multirow,booktabs,blindtext}
\usepackage{adjustbox}
\usepackage{subfig}
\usepackage{placeins} 
\usepackage{hyperref}
\usepackage{graphicx,rotating}
\usepackage{mwe}
\usepackage{txfonts}
\usepackage[T1]{fontenc}
\usepackage{epsfig}
\usepackage{natbib}
\bibpunct{(}{)}{;}{a}{}{,} 
\usepackage{color}
\usepackage{amssymb}  
\usepackage{tabularx}
\usepackage{tabulary}
\usepackage{comment}

\usepackage{ltablex}
\usepackage{caption, booktabs}

\DeclareGraphicsExtensions{.pdf,.png,.jpg}
\def\mgii{Mg\,{\sc ii}}
\def\civ{C\,{\sc iv}}

\def\hbeta{H$\beta$}
\def\mbh{$M_{\mathrm{BH}}$}

\begin{document} 

   \title{The Galaxy Activity, Torus, and Outflow Survey (GATOS). Black hole mass estimation using machine learning}

   \author{
    R. Poitevineau \inst{1,2}  
	 \and  
      F. Combes \inst{1,3}
      \and
      S. Garcia-Burillo \inst{4}
      \and
      D. Cornu \inst{1}
      \and
      A. Alonso Herrero \inst{5}
      \and
      C. Ramos Almeida \inst{6,7}
      \and
      A. Audibert \inst{6,7}
      \and
      E. Bellocchi \inst{8,9}
      \and
       P. G. Boorman \inst{10}
      \and
       A. J. Bunker \inst{11}
      \and
       R. Davies \inst{12}
      \and
       T. Díaz-Santos \inst{13,14}
      \and
       I. García-Bernete \inst{11}
      \and
       B. García-Lorenzo  \inst{6,7}
      \and
       O. González-Martín \inst{15}
      \and
       E. K. S. Hicks \inst{16}
      \and
       S. F. Hönig \inst{17}
      \and
       L. K. Hunt \inst{18}
      \and
       M. Imanishi \inst{19,20}
      \and
       M. Pereira-Santaella \inst{21}
      \and
       C. Ricci \inst{22,23}
      \and
       D. Rigopoulou \inst{11,14}
      \and
       D. J. Rosario \inst{24}
      \and
       D. Rouan \inst{25}
      \and
       M. Villar Martin\inst{4}
      \and
       M. Ward \inst{26}
     }

   \institute{Observatoire de Paris, LERMA, PSL University, Sorbonne Universit\'{e}, CNRS, F-75014, Paris, France  
            \and  
            Institute of Physics, Laboratory of Astrophysics, \'{E}cole Polytechnique F\'{e}d\'{e}rale de Lausanne (EPFL), 1290 Sauverny, Switzerland
            \and
            Coll\`{e}ge de France, 11 Place Marcelin Berthelot, 75231 Paris, France
            \and  
            Observatorio Astronomico Nacional (OAN-IGN)-Observatorio de Madrid, Alfonso XII, 3, 28014, Madrid, Spain
            \and 
            Centro de Astrobiología (CAB), CSIC-INTA, Camino Bajo del Castillo s/n, E-28692 Villanueva de la Ca\~{n}ada, Madrid, Spain
            \and 
            Instituto de Astrof\'{i}sica de Canarias (IAC), Calle V\'{i}a L\'{a}ctea, s/n, 38205 La Laguna, Tenerife, Spain
            \and 
            Departamento de Astrof\'{i}sica, Universidad de La Laguna, 38206 La Laguna, Tenerife, Spain
            \and 
            Departamento de F\'{i}sica de la Tierra y Astrof\'{i}sica, Fac. de CC Físicas, Universidad Complutense de Madrid, 28040 Madrid, Spain
            \and 
            Instituto de F\'{i}sica de Partículas y del Cosmos IPARCOS, Fac. CC Físicas, Universidad Complutense de Madrid, 28040 Madrid, Spain
            \and 
            Astronomical Institute, Academy of Sciences, B\^{o}cn\'{i} II 1401, 14131 Prague, Czechia
            \and 
            Department of Physics, University of Oxford, Oxford OX1 3RH, UK
            \and  
            Max-Planck-Institut f\"ur Extraterrestrische Physik, Postfach 1312, 85741 Garching, Germany
            \and 
            Institute of Astrophysics, Foundation for Research and Technology- Hellas, 71110 Heraklion, Greece
            \and 
            School of Sciences, European University Cyprus, Diogenes Street, Engomi, 1516 Nicosia, Cyprus
            \and 
            Instituto de Radioastronom\'{i}a y Astrofísica (IRyA), Universidad Nacional Autónoma de M\'{e}xico, Antigua Carretera a P\'{a}tzcuaro \#8701, Ex-Hda. San Jos\'{e} de la Huerta, Morelia, Michoac\'{a}n, M\'{e}xico C.P. 58089
            \and 
            Department of Physics \& Astronomy, University of Alaska Anchorage, Anchorage, AK 99508-4664, USA
            \and 
            School of Physics \& Astronomy, University of Southampton, Southampton SO17 1BJ, Hampshire, UK
            \and 
            INAF – Osservatorio Astrofisico di Arcetri, Largo Enrico Fermi 5, 50125 Firenze, Italy
            \and 
            National Astronomical Observatory of Japan, National Institutes of Natural Sciences (NINS), 2-21-1 Osawa, Mitaka, Tokyo 181-8588, Japan
            \and 
            Department of Astronomy, School of Science, Graduate University for Advanced Studies (SOKENDAI), Mitaka, Tokyo 181-8588, Japan
            \and 
            Instituto de F\'{i}sica Fundamental, CSIC, Calle Serrano 123, 28006 Madrid, Spain
            \and 
            Instituto de Estudios Astrof\'isicos, Facultad de Ingenier\'ia y Ciencias, Universidad Diego Portales, Av. Ej\'ercito Libertador 441, Santiago, Chile
            \and 
            Kavli Institute for Astronomy and Astrophysics, Peking University, Beijing 100871, China
            \and 
            School of Mathematics, Statistics, and Physics, Newcastle University, Newcastle upon Tyne NE1 7RU, UK
            \and 
            LESIA, Observatoire de Paris, PSL Research University, CNRS, Sorbonne Universit\'{e}, Paris-Cit\'{e} University 5 Place Jules Janssen, 92190 Meudon, France
            \and 
            Centre for Extragalactic Astronomy, Department of Physics, Durham University, South Road, Durham DH1 3LE, UK
            }

                \date{Accepted 23/11/2024}
   \titlerunning{GATOS: black hole masses}
   \authorrunning{R. Poitevineau et al.}

  \abstract
  {
  The detailed feeding and feedback mechanisms of Active Galactic Nuclei (AGN) are not yet well known, and for low-luminosity AGN, obscured AGN, and late-type galaxies, the masses of the central black holes (BH) are difficult to determine precisely. Our goal with the GATOS sample is to study the circum-nuclear regions, and in the present work, to better determine their BH mass, with more precise and accurate estimations than those obtained from scaling relations.
  We use the high spatial resolution of ALMA to resolve the CO(3-2) emission within $\sim$100~pc around the supermassive black hole (SMBH) of seven GATOS galaxies and try to estimate their BH mass when enough gas is present in the nuclear regions.
  We study the 7 bright ($L_{AGN}(14-150\mathrm{keV}) \geq 10^{42}\mathrm{erg/s}$), nearby (<28~Mpc) galaxies from the GATOS core sample. For the sake of comparison, we first searched the literature for previous BH mass estimations. We also made additional estimations using the \mbh~ - $\sigma$ relation and the fundamental plane of BH activity.  
  We developed a new method using supervised machine learning to estimate BH mass either from position-velocity diagrams or from first-moment maps computed from ALMA CO(3-2) observations.  
  We used numerical simulations with a large range of parameters to create the training, validation, and test sets.
  Seven galaxies have enough gas detected so that we can make a BH estimation from the ALMA data: NGC~4388, NGC~5506, NGC~5643, NGC~6300, NGC~7314, NGC~7465, and NGC~7582. Our BH masses range from  ~6.39 to 7.18~log$(M_{BH}/M_\odot)$ and are consistent with the previous estimations. 
  In addition, our machine learning method has the advantage of providing a robust estimation of errors with confidence intervals. The method has also more growth potential than scaling relations. This work represents the first step toward an automatized method for estimating \mbh~ using machine learning. 
  }
   \maketitle
%

\section{Introduction}
At the center of massive galaxies lie Super Massive Black Holes (SMBH), Black Holes (BH) with masses ranging from $10^6~M_\odot$ to $10^{10}~M_\odot$. Those BHs grow together with their host galaxies over time by accreting matter; the release of this considerable gravitational energy produces shocks and emits powerful radiation in the central region of the galaxies, a phenomenon called Active Galactic Nuclei (AGN).
There exists an empirical relation between the mass of the SMBH (\mbh~) and the bulge mass of their galaxy host, the latter often measured by its central velocity dispersion, the \mbh~ - $\sigma$ relation \citep[e.g.][]{kormendy_coevolution_2013,shankar_selection_2016,shankar2019,shankar2020,mardsen2020}. This suggests either common fueling mechanisms, and/or a feedback effect of AGN on their host. This scaling relation demonstrates a symbiosis and regulation from the BH, which transcends the merger and/or accretion history of the BH.

AGN feedback is often invoked to moderate or even stop star formation in their massive hosts when supernovae feedback is no longer efficient. Indeed, in the standard $\Lambda$CDM cosmological model, simulations of galaxy evolution in their dark matter halos over-predict the number of the most massive galaxies, with large stellar masses. Through halo abundance matching,
\cite{behroozi_average_2013} observe that around z$\sim$3 massive galaxies become less efficient at forming stars.
The comparison between the observed luminosity function and the simulated one for galaxies without feedback, reveals that star formation has not been efficient both at the low and high mass ends: galaxies have not been able to form as many stars as in the simulations. From the observed halo mass function, one can derive the stellar mass-to-halo mass ratio showing a peak at 20\% of the universal baryon to dark matter ratio, meaning that $>$80\% of the baryonic matter resides outside of galaxies. While at the low mass end, supernovae feedback can eject most of the baryons, only AGN feedback is efficient enough at the high mass end. 
While at high redshift, dense and cold gas can be accreted efficiently in dark matter halos, with still shallow potential wells \citep{Dekel2009}, at low redshift, gas is shock-heated when entering massive galaxies and deep potential wells, and only sparse cold gas is in falling. However, the hot gas could still cool down in the center of the structure, and feed efficiently star formation. Then AGN feedback is still necessary to moderate star formation \citep{Fabian_observational_2012}.

AGN feedback can be classified into two modes, the radiative (quasar) mode, and the kinetic (radio) mode. In the first case, luminous AGNs are powered by thin accretion disks. If the AGN bolometric luminosity is significant with respect to the Eddington luminosity, which is proportional to the BH mass, the radiation pressure creates a wind from the accretion disk. This is the quasar mode when the AGN luminosity is higher than 1\% Eddington \citep{Fabian_observational_2012}.
In the second case, for low luminosity AGN, the pressure and dragging of the relativistic jet and consequent shocks heat the gas, and a bubble/cocoon is inflated by the jet. This is the radio mode.

Estimating the SMBH mass is therefore fundamental, for a better understanding of their formation, in symbiosis with the galaxy growth. One of the best ways to measure its mass is to measure its gravitational influence on its surroundings, through the kinematics of the matter inside its Sphere of Influence. Another way is to rely on the empirically determined \mbh~ - $\sigma$ scaling relation but with more scatter at the low mass end \citep[e.g.,][]{ferrarese_fundamental_2000,gebhardt_relationship_2000}.

For nearby galaxies, it is possible to measure the BH mass (\mbh) from the stellar kinematics, as in M31 \citep{Bender2005}. The nuclear stellar disk, inside the Sphere of Influence of the black hole (SoI), is dominated by Keplerian rotation. The kinematics of central ionized gas can also be used for \mbh~ determinations. It offers multiple advantages. First, ionized gas emission lines are much easier to detect than stellar absorption lines, or even molecular lines. Secondly, the velocity dispersion of gas is much lower than that of stars, and it is easier to model its rotating Keplerian disk \citep[e.g.][]{kormendy_coevolution_2013}. 
However, gas dynamics has some major drawbacks. Enough gas should be detected to properly sample the nuclear dynamics, and it should be well distributed within the SoI of the BH. Gas is modeled as a collisional fluid, meaning that non-gravitational perturbations must be taken into account. Perturbations, such as outflows, can affect the gas kinematics locally and must be disentangled to obtain the \mbh. The true velocity dispersion must be distinguished from a beam-smeared rotational velocity gradient, to properly correct and obtain the circular velocity for the \mbh. This is why it is important to compare \mbh~ estimation made with ionized or molecular gas with masses obtained with other methods.

During the past few years, the domain of Machine Learning (ML) has widely expanded in the astrophysical community. It is a very powerful and flexible tool for data analysis that showed great performances on a wide variety of tasks using complex data. 
One of the main domains of the application of ML is image analysis. Convolutional Neural Networks (CNN) are efficient and powerful artificial neural network (NN) architectures to handle images. For example, they are used to classifying large numbers of galaxies \citep{Huertas2020}, in view of future missions like Euclid. They are also used to classify supernovae \citep{Lochner2016}, find strong lensing features \citep{Lanusse2018}, transient phenomena \citep{Mahabal2019}, photometric redshifts \citep{Pasquet2019} or star clusters \citep{Castro2019}. They have however not yet been used to estimate the mass of SMBH.

In this paper, we present a new ML method to estimate the mass of SMBH in galaxies. The masses obtained with this new method are compared with SMBH masses found in the literature and with other estimations made using the Fundamental Plane of BH activity (FPBH) and the \mbh~ - $\sigma$ relation. 
We describe  the galaxy sample and our selection criteria in Section 2. Section 3 focuses on the different methods to estimate the \mbh. We first introduce the \mbh~ - $\sigma$ and the reverberation mapping methods that were used to make previous estimations and calibrations. We also discuss the FPBH method used to complete the diverse masses from the literature.
Section 4 presents how we combine ALMA data cubes, numerical simulations, and a dedicated ML approach to estimate more precisely \mbh~ from the CO(3-2) gas dynamics. A large number of simulated models are used to evaluate the quality of our estimations.  
In Section 5, the results found with our ML method are presented. Finally, Section 6 and 7 summarize and discuss our findings and possible future improvements to our methodology.

\section{Sample}\label{sec:sample}

The GATOS international collaboration has for aims to study topics related to the physics taking place in the nuclear region of AGN. These encompass the gas flow cycle, the emission of polar dust, the properties of the torus/obscuring material, and the interplay between star formation activity and AGN phenomena. 
Here, we study the 10 galaxies from the GATOS core sample studied by \cite{garcia-burillo_galaxy_2021}. The core sample is composed of nearby galaxies ($<$28Mpc) and luminous nuclei, $L_{AGN}(14-150\mathrm{keV}) \geq 10^{42}\mathrm{erg/s}$. The upper limit on the distance is set to have a sufficient spatial resolution to study molecular tori as small as $\sim 10$pc in radius while the limit on the luminosity is set to avoid overlap with the ongoing ALMA surveys of nearby Seyferts such as the NUGA (Nuclei of Galaxies survey) \citep[e.g.][]{combes_alma_2019, audibert_alma_2019}. 

The present work uses the CO(3-2) data cubes from ALMA observations. All the details of the GATOS core sample and their observations are given in \cite{garcia-burillo_galaxy_2021}. We do not consider the three additional GATOS galaxies NGC~1068 \citep[e.g.][]{garcia-burillo_alma_2019}, NGC~1365 \citep[e.g.][]{combes_alma_2019} and NGC~3227 \citep[e.g.][]{alonso-herrero_nuclear_2019}.

Not all 10 data cubes are appropriate for the determination of the central kinematics and \mbh~ estimation. Two of the galaxies, NGC~6814 and NGC~7213 do not present enough CO emission in the very center; their central pixels are empty or with some emission with an S/N ratio below 2. For another galaxy, moment maps are not usable for a \mbh~ estimation with the ML approach; NGC~4941 has too little gas detected, and its distribution is too sparse.  
There remain therefore 7 galaxies: NGC~4388, NGC~5506, NGC~5643, NGC~6300, NGC~7314, NGC~7465 and NGC~7582.

%
\section{Classical methods for estimating \texorpdfstring{\mbh}{}}

    \subsection{Dynamical estimation}\label{sect:dynamical mbh}
The estimation of SMBH mass through dynamical models involves studying the movements of celestial objects, like stars or gas, around the SMBH. These observed motions, often referred to as tracers, serve as constraints for the dynamical models. These models discern the contribution of the BH to the potential from that of the surrounding galaxy, thus facilitating the inference of the SMBH mass by fitting these models to observations \citep{kormendy1995,ferrarese_fundamental_2000,kormendy2004,gebhardt_relationship_2000,kormendy_coevolution_2013}.

A notable constraint of this methodology is its reliance on resolving the kinematics of objects situated near the SMBH. Consequently, it becomes impractical for galaxies located at considerable distances. Therefore, alternative methodologies are required to estimate SMBH masses in such distant galaxies.

Across various dynamical models, an essential consideration is the necessity for a realistic representation of the gravitational potential of the galaxy. Additionally, the data characterizing the tracers motions must be robust enough to detect the presence of a BH \citep{gebhardt_axisymmetric_2003,Ferrarese2005,kormendy_coevolution_2013}. Among the multitude of intricate processes connected to SMBHs and their immediate surroundings, the study of stellar phenomena proves more amenable to analysis and modeling. This feasibility arises from stars often being treatable as point masses, predominantly influenced by gravity. The substantial mass disparity between stars and \mbh~ simplifies the problem further, allowing stars to be considered test masses on intermediate scales. Here, they operate within the \mbh~ potential, where gravitational interactions among stars are negligible, yet they are distant enough to avoid significant tidal effects from the \mbh.\\

The molecular gas surrounding the SMBH serves as another dynamical tracer for estimating the \mbh. Gas-dynamical mass measurements present distinct advantages compared to stellar-dynamical modeling. The simplicity inherent in gas-dynamical modeling stems from the gas Keplerian rotation within a dynamically cold disk. Unlike the intricate orbit-based computations necessary for treating stars, analyzing gas dynamics in a rotating disk proves computationally less demanding. The reduced complexity allows neglecting factors such as orbital anisotropy, triaxiality, or the influence of the dark matter halo. 
The gas-dynamical approach assumes a thin, rotating disk following circular orbits within the principal plane of the galaxy potential \citep{marconi2001,marconi2003}. This methodology aims to calculate a model velocity field aligning with observed velocities, velocity dispersions, and the line emission's surface brightness distribution. The gravitational potential of the galaxy comprises contributions from stars, measured by the projected stellar surface brightness, and assumes a mass-to-light ratio along with the presence of a BH \citep{melchior2011,combes_alma_2019}.

However, gas dynamics introduce several complications \citep{kormendy_coevolution_2013}. Properly sampling the BH SOI necessitates distributing gas across radii. Simultaneously, the gas kinematics must display sufficient orderliness for interpretation. Furthermore, gas differs from stars as it behaves like a collisional fluid, responsive to non-gravitational disturbances such as turbulence, shocks, radiation pressure, and magnetic fields. 
Another challenge in gas dynamics involves dust absorption, potentially rendering the gas distribution opaque. Consequently, assumptions about observing through the gas in projection may not hold under such circumstances. Each galaxy requires meticulous examination to confirm whether the gas has reached an equilibrium configuration, influenced mainly by gravitational effects, considering the opacity caused by dust absorption.

\begin{figure}
    \hspace*{-0.5cm}
    \captionsetup[subfigure]{labelformat=empty}
        \includegraphics[width=1.0\linewidth,height=30cm,keepaspectratio]{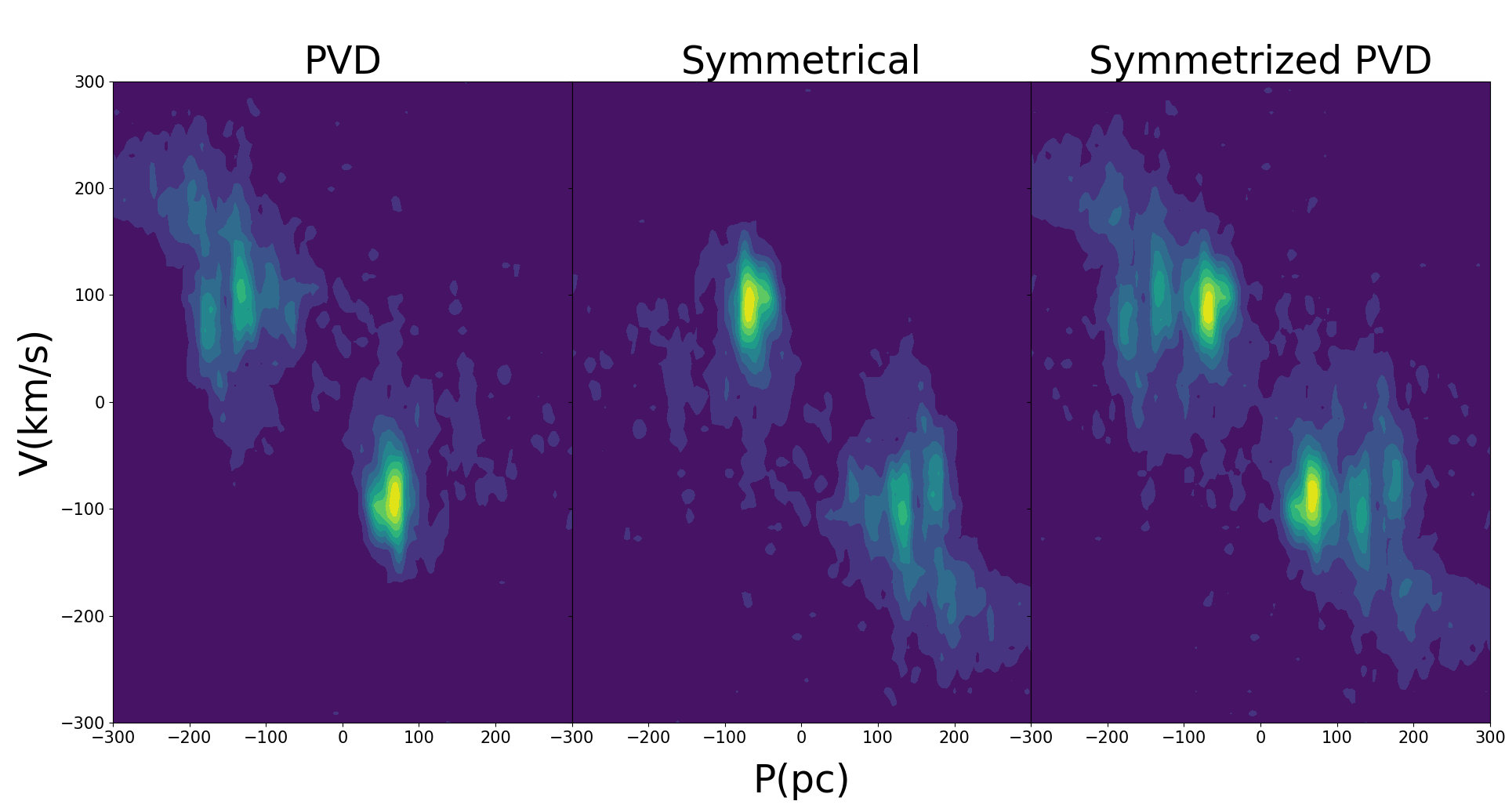}
     \caption{Example of NGC~4388 symmetrized PVD}
    \label{fig:sympvd4388}
\end{figure}

To investigate the gas dynamics of CO(3-2), a useful source of information is the Position-Velocity diagram (PVD) obtained along the major axis, which is widely employed for this purpose, for instance by the WISDOM project \citep[e.g.][]{onishi_wisdom_2017, davis_wisdom_2017,davis_wisdom_2018}. 

To estimate the \mbh~ from the PVD, we fit the sum of the contributions from the various galaxy central components: bulge, BH, disk, and possible bar and stellar nucleus. 
\begin{table}[h]
\caption{\label{table:contrib}Stellar mass components}
\vspace{-0cm}
\hspace*{-0.0cm}
\renewcommand{\arraystretch}{1.125}
	\begin{tabulary}{\linewidth}{cccccc}
		\toprule
		 ID & Comp & $R_*$(kpc) & $M_{*}(10^{10}M_{\odot})$ & Sersic n & B/T \\
		\midrule  
		NGC~4388   & Disk    & 1.68  & 2.44 & 1.0   & 0.84 \\
		        & Nucleus &       & 0.46 &       &       \\
		  \hline
		NGC~4941   & Bulge   & 0.67  & 0.50 & 11.5  & 0.24 \\
		        & Disk    & 1.13  & 0.95 & 1.0   &       \\
		        & Bar     & 10.92 & 0.66 &       &       \\
		        & Nucleus &       & 0.01 &       &       \\
		  \hline
		NGC~5643   & Bulge   & 0.20  & 0.89 & 2.9   & 0.13 \\
		        & Disk    & 9.12  & 4.61 & 1.0   &       \\
		        & Bar     & 3.75  & 1.17 &       &       \\
		        & Nucleus &       & 0.01 &       &       \\
		  \hline
		NGC~6300   & Bulge   & 4.07  & 0.875 & 1.7   & 0.19 \\
		        & Disk    & 44.11 & 1.375 & 1.0   &       \\
		        & Bar     & 27.80 & 2.240 &       &       \\
		        & Nucleus &     & 10$^{-4}$ &     &       \\
		  \hline
		NGC~7314   & Disk    & 1.70  & 2.105 & 1.0   & 0.0 \\
		        & Nucleus &       & 0.012 &       &       \\
		  \hline
		NGC~7465   & Bulge   & 0,34  & 0.486 & 1.0   & 0.19 \\
		        & Disk    & 2.51  & 0.630 & 1.0   &       \\
		        & Bar     & 1.35  & 1.124 &       &       \\
		  \hline
		NGC~7582   & Bulge   & 0.19  & 1.132 & 3.7   & 0.35 \\
		        & Disk    & 2.24  & 1.989 & 1.0   &       \\
		        & Bar     & 12.54 & 0.135 &       &       \\
		
		\bottomrule
	\end{tabulary}
    \tablefoot{Stellar mass components from the \cite{salo_spitzer_2015} Galfit decomposition from $S^4G$ $3.6\mathrm{\mu m}$ except for NGC5643 and NGC6300 where the Galfit decomposition is our own fit from the WCF3 Hubble Legacy Archive image.
    } 
\end{table}

To model the stellar distribution in each galaxy, we use the \cite{salo_spitzer_2015} Galfit \citep{peng2010AJ....139.2097P} decomposition from the Spitzer Survey of Stellar Structure in Galaxies $3.6 \mathrm{\mu m}$ infrared images \citep{spitzer2015ApJS..219....3M}. Two of our galaxies, NGC~5643 and NGC~6300, were not in the \cite{salo_spitzer_2015} sample, so we made the Galfit decomposition ourselves. The physical characteristics of the different galaxy components of our sample are shown in Table~\ref{table:contrib}.

The models employed to fit the galaxies exhibit symmetry, while the observed PVD does not. To address this asymmetry, we adopt a symmetrization approach by generating a symmetrical version of the observed PVD by mirroring it about its central pixel, followed by the summation of the original and mirrored versions. An illustration of the process employed to obtain the symmetrized PVD of NGC~4388 is provided in Fig.\ref{fig:sympvd4388}. Since the symmetrization tends to keep the highest gradients on the PVD, we can expect that the \mbh~ estimations might be biased high.

In a Keplerian rotating disk, the gas rotation velocity is inversely proportional to the square root of the radius. When the gas is just entering the SoI of the BH, the main contribution to the velocity is the BH, but there is also some contribution from the bulge stars. Thus, the measured velocity, corrected from velocity dispersion, can put an upper limit on the \mbh. 
If no gas is observed in the SoI, an estimation can be done from the closest gas detected instead, assuming an accurate estimation of the bulge mass. Large error bars can be expected if the gas is far from the SoI, and/or if the bulge mass is not precisely known.

    \subsection{\texorpdfstring{\mbh~ - $\sigma$}{} relation}
The tight correlation between BH and bulge masses drawn from the \mbh~ -$\sigma$ relation \citep[e.g.][]{ferrarese_fundamental_2000, gebhardt_relationship_2000} suggests that there is a mechanical and/or radiative feedback from the AGN on the star formation in galaxies, all along their growth through accretion and mergers. The origin of this coupling is not yet fully understood, but there are analytic and numerical models created to constrain the physical mechanism and reproduce the \mbh~ - $\sigma$ relation \citep[e.g.][]{silk_quasars_1998,granato_physical_2004}. Nevertheless, the observational \mbh~ - $\sigma$ relation is not well constrained for low-mass galaxies ($M<10^9~M_\odot$). The tight correlation is mostly derived from biased galaxy samples with a high \mbh, which have a larger SoI easier to resolve with common angular resolutions \citep[e.g.][]{silk_quasars_1998, costa_feedback_2014}. In this work, we used the \cite{kormendy_coevolution_2013} correction of the \mbh~ - $\sigma$ relation:

\begin{equation}
    \mathrm{log} M_{BH}(M_{\odot}) = 8.5 \pm 0.05 + (4.41 \pm 0.29) \mathrm{log}\left(\frac{\sigma}{\mathrm{200km/s}}\right),
\end{equation}
where $\sigma$ is the stellar central velocity dispersion. For our sample, it was taken from the Hyperleda compilation \citep[e.g.][]{makarov_hyperleda_2014}. This relation has an intrinsic error of 0.3 dex.

\subsection{Reverberation mapping}
BLRs have light days to light weeks radii, which are too small to be spatially resolved. However, they can be resolved in time with the Reverberation Mapping technique \citep{blandford_hydromagnetic_1982,bahcall_time_1972,lyutyi_narrow-band_1972,cherepashchuk_rapid_1973}. The SMBH accretion disk shocks emit UV radiation that heats the BLR gas clouds, ionizing them and exciting emission lines. Thus, when the UV continuum source varies, the emission lines vary too ('reverberate') but with a delay of $R_{BLR}/c$. Therefore, the BLR radius can be estimated by measuring the time delay between the change in emission from the central source and its impact on the BLR lines. Then, assuming that the width of the BLR emission lines is due to the gravitational influence of the central SMBH, we can derive the following relation from the Virial theorem:

\begin{equation}\label{eq:RM}
    M_{\mathrm{BH}} = \frac{f~(\Delta V^2)R}{G}
\end{equation}
where \mbh~ is the mass of the central BH, $G$ is the gravitational constant, $\Delta V$ is the virial velocity of the BLR, $R$ is the BLR radius and $f$ is a factor depending on the BLR geometry. The fact that the BLR velocity is dominated by the BH mass (virial velocity and gravitational redshift), and not by possible gas outflows, was validated by \cite{gaskell_direct_1988}.

Most of the uncertainties of this method come from the factor $f$. \cite{kollatschny_broad-line_2011} observed that the BLR structure changes with the emission-line turbulent dispersion and rotational width, implying that f depends on these two factors. 
The findings of \cite{gaskell2008RMxAC..32....1G,gaskell2011BaltA..20..392G} indicate another inherent constraint, which suggests that the continua of AGN can exhibit localized flares in off-center regions situated relatively close to the inner boundary of the BLR radius. 

Even in the absence of the expensive and time-consuming monitoring observations, it is possible to use this method, thanks to the AGN-Luminosity vs BLR-size scaling relation, calibrated from reverberation mapping measurements \citep{Wandel1999}. Refining the method with a large reverberation mapping database, and comparing it to the other methods, \cite{peterson_central_2004} find that the precision of the method is $\sim$ 30\%, comparable to the gas or stellar kinematical method. 

This can be explained by considering that the reverberation mapping technique is notably influenced by the inclination angle of the BLR region, contrary to the gravitational redshift method. As demonstrated in the study by \cite{liu2017}, the substantial disparity between gravitational and virial mass estimates can be reconciled by adjusting the parameter $f$ to a value between 8 and 16 for the virial measurement. This adjustment of $f$ depends on radiation pressure and accretion rate, in addition to the inclination of the BLR. The conventional value of $f\sim1$ corresponds to an inclination angle of around 30 degrees.
    \subsection{Single epoch method}\label{sec:single_epoch}
The single epoch method uses the relation between the BLR size and the AGN optical/UV continuum luminosity empirically found from reverberation mapping \citep{peterson_central_2004,kaspi_reverberation_2007,bentz_lick_2009}, as well as the tight correlation between the continuum luminosity and that of broad emission lines \citep[e.g.,][]{shen_catalog_2011}. 
With these considerations, the \mbh~ can be expressed as

\begin{equation}
\label{eq:MBH}
\log \left( \frac{M_{\rm BH}}{M_{\odot}} \right) = a + b\;\log \left( \frac{L}{10^{44}~{\rm erg~s}^{-1}} \right) + c\;\log \left( \frac{\rm FWHM}{{\rm km~s}^{-1}} \right)\;,
\end{equation} 

where the coefficients $a$, $b$, and $c$ are empirically calibrated against local AGN with reverberation mapping masses or using different lines. $L$ and FWHM are the line luminosity and width. 
At low redshift, below 0.75, \hbeta~ is the most commonly used line.  For redshift between 0.75 and 2, the \mgii~ $\lambda 2800$ line is a common choice \citep{mclure2002}. And for $z \gtrsim 2$, the best line to use is \civ~ $\lambda 1549$ \cite{vestergaard_determining_2002}.
Since their first estimation, the relation for these lines had been revised by
 \cite{vestergaard_determining_2006,wang2009,shen_catalog_2011,xiao2011} and others.

The single epoch method has the advantage of being inexpensive in telescope times since it can estimate the SMBH mass from just a spectrum. However, it is a more indirect method than reverberation mapping and does not contain as many indicators that the measured velocities are virialized ones. This method has an uncertainty of $\sim 0.3$ dex.

    \subsection{Fundamental plane of BH activity} 
The FPBH is a strong correlation between the \mbh, its 5 GHz radio continuum luminosity, and its 2-10~keV X-ray power-law continuum \citep{merloni_fundamental_2003}. When accretion disks go through episodes of low-luminosity advection-dominated accretion flow, they are observed to be in a hard state (hard X-ray radiation). In this state, they are often accompanied by relativistic jets \citep{gallo_universal_2003, narayan_low-luminosity_2005}. During this state of inefficient radiation, the accreting material stops flowing in a disk at some distance from the accreting object. This new configuration results in the creation of a hot, optically-thin gas corona around the BH. Photons, from the now truncated disk, are Compton-up scattered to tens to hundreds of keV when they traverse this corona. Furthermore, the synchrotron mechanism in the relativistic jets produces emission primarily in the radio band \citep{blandford_models_1984}. The entire broadband combined emissions of the accretion disk, corona, and relativistic jets range from the radio to the X-ray bands. This emission can also be dominated by the jet itself, particularly if accretion rates are higher than $\sim 0.01 \dot{M}_{\mathrm{{Edd}}}$ \citep{falcke_scheme_2004}. 
A fainter thermal component emanating from the shortened accretion disk may also be identifiable alongside the primary non-thermal corona emission.
The jet is linked to the accretion process, and it has been shown that the luminosity of the X-ray and radio emissions are correlated and the disk-jet mechanism is independent of the \mbh. By considering these arguments, \cite{merloni_fundamental_2003} and \cite{falcke_scheme_2004} probed a large sample of galactic BH and SMBH and found a strong correlation between the X-ray luminosity (between 2 and 10~keV), the radio luminosity (5~GHz) and the \mbh. This correlation is called the fundamental plane of BH activity.
According to \cite{gultekin_fundamental_2019} the \mbh~ can be estimated with the relation:

\begin{equation}
\begin{split}
    \mathrm{log}\left(\frac{M_{\mathrm{BH}}}{10^{8}M_{\odot}}\right) = 0.55 &\pm 0.22 + (0.19 \pm 0.10) \mathrm{log}\left(\frac{L_{R}}{10^{38}\mathrm{erg/s}}\right) \\
    &+ (-0.59^{+0.16}_{-0.15}) \mathrm{log}\left(\frac{L_{x}}{10^{40}\mathrm{erg/s}}\right),
\end{split}
\end{equation}
where $L_{x}$ is the 2-10 keV X-ray power-law continuum and  $L_{R}$ the 5~GHz radio luminosity calculated as $L_{R} = \nu L_{\nu} = (\nu=5 \, \mathrm{GHz})\times 4\pi D^2 F_{5}$, $F_{5}$ being the 5~GHz radio flux density. This relation has an intrinsic error of $\sim$0.40 dex.

This FPBH scaling relation is a useful method to distinguish between X-ray binaries, intermediate mass BH, and SMBH, and to determine the \mbh~ in a Type 2 AGN, or in a host galaxy with disturbed morphology, where the \mbh -$\sigma$ method is unusable. The only requirement to estimate the \mbh~ is to have X-ray and radio data of sufficient quality to resolve the central source and also confirm its spectral characteristics (i.e. that it is indeed in the hard state). The data must have a high angular resolution to avoid contamination from other sources and the host. This is especially important to distinguish low luminosity AGN from X-ray binaries.

\section{ML approach to estimate \texorpdfstring{\mbh}{}}

We have chosen to employ ML techniques to estimate the BH mass using the ALMA data cubes of the GATOS core sample. 
These 3D data cubes can be effectively transformed into 2D images, enabling us to apply more easily CNNs for analysis. 
Although everyday data tend to be more complex, CNNs have proven to be highly effective in such cases, leading us to expect their suitability for astrophysical data as well. 
This study aims to serve as a proof of concept that ML techniques can be used to infer \mbh~ from observational data.

As of now, our sample consists of only 7 galaxies, which implies that applying an ML method to such a small dataset may not yield the most optimal results. However, we anticipate the availability of a larger sample in the future. Additionally, the presence of data in the ALMA archives presents further opportunities for analysis and investigation, thus motivating our efforts toward automating the BH mass estimation process. 
Although scaling relations offer a certain level of automation, their intrinsic errors restrict the full exploitation of the high-resolution data. 
The ML method presented in this study serves as a proof of concept that ML can be applied to ALMA data to estimate \mbh~ with physical significance and is a first step toward an automated approach that leverages the data resolution to achieve precise \mbh~ estimations.

\subsection{Supervised learning method}

An artificial NN can be seen as a sequence of mathematical operations, organized in layers and composed of adjustable weights, that map an input to an output. 
In a supervised learning method, a dataset of inputs with known outputs, called the training set, is used to parameterize the weights so the NN infers the mapping function. It is done by feeding the training set to the NN and comparing its outputs to the expected ones. 
A "loss function" is used to compare the NN output to the expected one. 
The loss function serves as a measure of how good the NN is at predicting the expected outcome. 
The more accurate the output, the smaller the loss function. The goal of the training phase is to find a set of weights that minimize the loss function. To do so, the loss function is computed for each input in the training dataset. Then, the gradient of the loss function is used to slightly modify the weights via a backpropagation algorithm. This process is repeated for each input in the training data set. One pass of all the training data through the NN is called an epoch.

Overfitting is a common problem when training an artificial NN with a supervised ML method, and alerts when the NN infers a mapping function that is too specific to the training set, making it unable to perform accurately against unseen data. 
To avoid falling into such a regime, another dataset, called the validation set, is used. The validation set has the same statistic distribution as the training set but is never used to update the network weights. 
During the training, the loss function values of the training set and the validation set are regularly monitored to ensure that the network is still learning a generalized mapping function. 
Finally, after the training process is done, a third dataset, the test set, is used to estimate the performance of the NN on unseen data.  This practice aids in preventing the training process from prematurely halting upon an overestimation of model performance on the validation set, which may occur due to limited sample sizes or biased sample selection.

The principle of our approach is to run a large number of numerical simulations, fitted to each studied galaxy. With this method, the gravitational potential of the galaxy disk is assumed to be known from HST images, within some uncertainties. 
The simulations are subsequently conducted by systematically varying the parameters within the defined uncertainty ranges. This includes the mass and scales of the bulge (plus possible nucleus) and disk (plus possible bar) components, as well as the characteristics of the gas and stellar distributions (notably Sersic index).  
Inclination and position angles are fixed to the values estimated from the observations \citep{garcia-burillo_galaxy_2021}, except in some specified cases, where the nuclear disk is tilted with respect to the large-scale disk. The possible mass of the SMBH is varied in a wide range, compatible with the observations. 
Numerous numerical simulations with different \mbh~ are used to train an artificial NN model for each galaxy. After the training is done on the simulations, the trained model can then be used to infer the estimated \mbh~ from the real observation.

The sample size of seven galaxies is insufficient for training a NN to generalize across different galaxies due to the high variability and complexity in their physical properties. A larger and more diverse dataset is essential to capture the wide range of conditions present in the universe, which is not feasible with just seven galaxies.
However, this limited sample size makes it practical to train a separate NN for each galaxy. By focusing on individual galaxies, we simplify the problem and can determine if our methodology works effectively on a smaller scale. Training separate networks for each galaxy involves creating distinct training, validation, and test sets for each one. For each observed galaxy, we generated 25,000 simulated data cubes, dividing them into a training set of 16,500 simulations, a validation set of 3,500 simulations, and a test set of 5,000 simulations.
This approach allows us to study the behavior of the networks more precisely, as we can evaluate their performance in a controlled and consistent manner. By doing so, we can identify any potential issues or improvements needed in our model before attempting to generalize the approach to a more extensive and diverse set of galaxies. If successful, this method can serve as a foundation for scaling up to a broader application, ultimately aiming to develop a generalized model capable of accurately inferring \mbh~ across various galaxies.

\subsection{Numerical simulations}\label{subsection:numerical_simulation}

To simulate our data cubes, we used the model described in \cite{melchior2011} and \cite{combes_alma_2019}. 
We simulate gas particles in a potential that depends on the physical parameters of the galaxies and their mass components. Since we simulate only the nuclear disk with a radius of typically $\sim 100$~pc, the dark matter contribution to the gravitational potential is negligible. 
Our models are axisymmetric homogeneous gas disks \citep[e.g.][]{miyamoto_three-dimensional_1975} with gas particles on nearly circular orbits, with low-velocity dispersion, corresponding to thin cold disks. The sizes of the modelled disks are taken from the observed CO(3-2) disks. 
The velocity distribution of the gas particles is set such that the Toomre Q parameter of the disk is equal to one and the ratio between tangential and radial velocity dispersion is taken from the epicyclic theory \citep[e.g.][]{toomre_gravitational_1964}.
To ensure sufficient statistics, we fixed the number of particles at $10^6$.

To compare the model with observational data, we generated data cubes by projecting the model onto the sky using the best-fit large-scale inclinations and position angles. We then computed the line-of-sight velocity distribution. The pixel size in the data cubes was chosen to match that of the observed data (ranging from 4 to 7 pc, depending on the galaxy), with velocity channels of 10 km/s. The data were smoothed to match the observed beam, resulting in a cube of dimensions (250, 250, 60) for each simulation. 
Since the gas distribution exhibits asymmetry and patchiness, which affect the mass-weighted velocity within each observed beam, we applied a normalization process to the model cube. Specifically, we normalized the model cube pixel by pixel in the 2D projection using the zeroth moment map of the CO observations. This normalization serves as a multiplicative filter for our homogeneous gas disks. Consequently, each CO spectrum at every position in the model is normalized to the observed integrated flux at that corresponding position.

The different parameters of the simulations are: the mass, radius, and height of the galaxy bulge ($M_{\mathrm{bulb}}$, $R_{\mathrm{bulb}}$, $H_{\mathrm{bulb}}$); the corresponding properties of the stellar disks ($M_{\mathrm{gal}}$, $R_{0}$, $H_{t}$); the gas-to-stellar mass ratio within the stellar disk ($RAPM$); the gas disk radius, height, inclination angle, and position angle ($R_{\mathrm{gas}}$,$H_{\mathrm{gas}},$$IA$,$PA$) and the SMBH mass (\mbh).

In the case of NGC~5643, it is observed that the $PA$ changes at approximately 30~pc. To model this behavior, we introduced a second $PA$ that replaces the initial one after the 30~pc mark.


For each simulation, all the parameters are taken randomly from a uniform distribution between the limits with physical meaning (see Table~\ref{table:contrib}), except the $IA$ and the $PA$. We can deduce from the moment maps a range for the radial scale of the gas $R_{\mathrm{gas}}$ and since the molecular disk is thin, we choose to keep the ratio $H_{\mathrm{gas}}/R_{\mathrm{gas}}<0.1$. For the stellar disk $R_{0}$ and $M_{\mathrm{gal}}$ are large with respect to the scale probed by our observations/simulations, and their possible variations are small. 
The $IA$ and $PA$ are set constant in all the simulations for a galaxy. They were taken as estimated by \cite{garcia-burillo_galaxy_2021} based on the gas kinematics.


\subsection{Preparing the inputs}

The GATOS CO(3-2) molecular gas observations are in the form of spectroscopic data cubes which are three-dimensional arrays that combine spatial and spectral information. Two of the dimensions represent the spatial coordinates and define the position in the sky. The third dimension represents the intensity or flux of light at each spatial position and each wavelength or frequency.

Gathering information directly from spectroscopic data cubes can be a challenging task, especially when these data cubes are utilized as inputs in ML processes \citep{ska2023}. The introduction of a third dimension into the input data significantly amplifies data complexity and can lead to considerable increases in computational requirements. To mitigate the resulting computational load and efficiently extract valuable insights from the data cube, it becomes essential to optimize and fine-tune the ML architectures employed.

However, we can take an easier approach and pre-process the data cubes to put the meaningful information more straightforwardly in the shape of two-dimensional images.  
This will make it easier for the NN to learn how to estimate the \mbh~ from the simulated data cubes. 
Two commonly used representations to study the CO(3-2) gas dynamics are the PVD and the moment maps (see Fig.\ref{fig:NGC7582_mom}).

The PVD allows the gas velocity and concentration as a function of the radius in a chosen direction to be seen in detail. Since the PVD is used to better see the effect of the SMBH on the gas dynamics, we select the PVD along the axis with the higher velocity gradient at the smallest radius possible.
Nevertheless, this representation has certain limitations. The PVD solely provides insight into a particular direction of the gas dynamics, thereby leading to the partial depiction of information. Consequently, the selection of a specific direction can influence the bias in the information presented in the PVD. Furthermore, studying the gas in only one direction makes the PVD more sensitive to a lack of gas at a specific radius, such as the center of NGC~4388 (\ref{fig:NGC4388_mom}).

In addition to studying the gas in one specific diagram, we consider a representation that encapsulates more information about the entire gas dynamic distribution by calculating the first two moment maps:\\
The 0th-moment map is the integrated intensity over the spectral axis, proportional to the gas surface density.
\begin{equation}
    M_0 = \int I_{\nu} d\nu
\end{equation}
The 1st moment represents the velocity $V$ weighted by the intensity.
\begin{equation}
    M_1 =  \frac{\int V  I_{\nu} d\nu}{M_0} 
\end{equation}

The PVD and the first moment maps are calculated for each simulated cube. Finally, moment by moment, and for the PVD, we normalize the value of each pixel by dividing it by the highest one. The PVD and first-moment maps of the galaxies in our sample are presented in Appendix \ref{app:contmaps}. 
Initially, the estimation of the SMBH mass will be conducted using the PVD as input, followed by employing the first-moment map as input.
\FloatBarrier

\subsection{ML architecture}

\begin{table}[h]
\caption{\label{table:archi_pv}NN architecture for \mbh~ estimations with PVD}
\begin{center}
	\begin{tabulary}{\linewidth}{cccc}
		\toprule
		 Layer & Filters & Size & Stride\\
		\midrule  
		 Conv1       & 16 & 7x7 & 2\\
		 Conv2       & 20 & 7x7 & 1\\
		 Conv3       & 20 & 7x7 & 2\\
		 Conv4       & 24 & 7x7 & 1\\
		 Conv5       & 64 & 5x5 & 2\\
		 Conv6       & 80 & 5x5 & 1\\
		 Conv7       & 80 & 5x5 & 2\\
      Conv8       & 128& 5x5 & 1\\
		 Max Pooling &    & 3x3 & 2\\
		 
		 Dense1 & 2048 &  & \\
		 Dense2 & 1024 &  & \\
		 Dense3 & 512  &  & \\
		 Dense4 & 1    &  & \\
		 
		\bottomrule
	\end{tabulary}
\end{center}
\end{table}

\begin{table}[h]
\caption{\label{table:archi_mom}NN architecture for \mbh~ estimations with moment maps}
\begin{center}
	\begin{tabulary}{\linewidth}{cccc}
		\toprule
		 Layer & Filters & Size & Stride\\
		\midrule  
		 Conv1         & 32 & 11x11 & 4\\
		 Max Pooling 1 &    & 3x3   & 2\\
		 
		 Conv2         & 64 & 5x5 & 1\\
		 Max Pooling 2 &    & 3x3 & 2\\
		 Conv3         & 64 & 5x5 & 1\\
		 Max Pooling 3 &    & 3x3 & 2\\
		 Conv4         & 64 & 5x5 & 1\\
		 Max Pooling 4 &    & 3x3 & 2\\
		 
		 Conv5         & 96 & 3x3 & 1\\
		 Conv6         & 96 & 3x3 & 1\\
		 Conv7         & 96 & 3x3 & 1\\
		 Max Pooling 5 &    & 3x3 & 2\\
		 
		 Dense1 & 2048 &  & \\
		 Dense2 & 1024 &  & \\
		 Dense3 & 512  &  & \\
		 Dense4 & 1    &  & \\
		 
		\bottomrule
	\end{tabulary}
\end{center}
\end{table}
To make our \mbh~ estimations via ML we used a CNN-supervised training method \citep[e.g.,][]{lecun_deep_2015}. 
Our NNs were developed using Python's library Keras \citep{chollet2015keras}. 
The PVD and the moment maps represent the information differently, so to get the best out of their distinct characteristics we need different CNN architectures. 
Thus, We will use two different architectures to make the \mbh~ estimations, one taking the PVD as input and one taking the first moment maps.

The NN taking the PVD as inputs is composed of a succession of eight convolution layers, a max pooling layer, and at the end, four dense layers (i.e. Table~\ref{table:archi_pv}). 
The NN that takes the first moment maps as inputs is composed of a succession of convolution and max pooling layers followed by four dense layers (i.e. Table~\ref{table:archi_mom}).
In both of these architectures, the fourth dense layers return the \mbh.

Dropout was added to the second dense layer in both architectures with a drop rate of 20\%. It means that in this layer, 80\% of the neurons are randomly selected to remain active at each path through the network. Usually, the dropout technique is used as a regularization method during training to help prevent overfitting \citep{srivastava_dropout_2014}. It can also improve the predictive performance of NNs in several tasks. 
Another way to use dropout is to keep it activated at inference time. 
At each inference, the dropout will randomly select a different combination of neurons, resulting in a different estimation. 
It is equivalent to an approximation of the probabilistic deep Gaussian process. 
Thus, making multiple inferences for the same input will not give one value but a probability distribution on the predicted value, similar to what is produced by a Monte Carlo Markov Chain process. This distribution can be studied to characterize the model uncertainties for that input.

We used the Root Mean Square Error function (RMSE) as loss function. It is a common loss function used in regression problems:
\begin{equation}
    RMSE = \sqrt{ \frac{\sum^{n}_{i=1}(Y_{pred,i} - Y_{real,i})^{2}}{n} }
\end{equation}
with $n$ the number of predictions made, $Y_{pred,i}$ the predicted values and $Y_{real,i}$ the actual values. To help stabilize the weights in our models during the training, we normalized the output targets (i.e., the \mbh) by dividing them by the highest value among the output targets.

We trained our models during 2000 epochs and took the weights of the epoch with the lowest RMSE on the validation set. We used an initial learning rate of 0.0001 and the Adam optimizer \citep{adamopt}. The training was made with a mini-batch size of 128 images.

\subsection{Estimating the reliability of models} \label{subsection:stats}

Multiple statistical indicators can be used to evaluate the quality of a regression model. These indicators can be used as metrics to help compare the training results of the different models. 
A first indication of the quality can be obtained by looking at the value of the loss function on the test set. 
However, studying the loss function alone is not enough to evaluate a model because it can be highly degenerate. Furthermore, the RMSE function goes from 0 (best fit possible) to $+\infty$ (worst fit possible). The lack of an upper bound can make the RMSE interpretation difficult but in our case, the error directly measures the average difference between the estimated \mbh~ and the actual ones.

A more reliable indicator would be the coefficient of determination \citep{Wright1921CorrelationAndCausation}\footnote{https://naldc.nal.usda.gov/download/IND43966364/PDF}. It is a statistical value that represents the proportion of a dependent variable variance that is predictable from the independent variables:
\begin{equation}
  R^{2} = 1 - \frac{\sum^{n}_{i=1}(Y_{pred,i}-Y_{real,i})^{2}}{\sum^{n}_{i=1}(\overline{Y_{pred}}-Y_{real,i})^{2}}
\end{equation}
where the notations are the same as the RMSE ones and $\overline{Y_{pred}}$ is the mean of the predictions. In a regression problem, the coefficient of determination measures how well the model fits the real data. It has the advantage of being confined by construction between 0 and 1, 0 being the worst fit possible and 1 meaning that the model perfectly fits the data. 
By studying the RMSE and the $R^2$ we can have a global evaluation of the capacity of a model to make the correct prediction.

To gain more information about the in-depth behavior of a model, we also use the relative errors (RE):
\begin{equation}
    RE_{i} = \frac{Y_{pred,i}-Y_{real,i}}{Y_{real,i}}
\end{equation}
Studying the distribution of the RE of a model when predicting the test set can reveal any systematic bias during the predictions. Furthermore, since our datasets are made of simulations, we can check if any of the simulation parameters primarily influence the results of the predictions by looking at the relations between the parameter values and the RE.

\section{Results on \texorpdfstring{\mbh}{} estimation}
\subsection{Reliability of ML models}

As described above in Sec \ref{sec:sample}, 7 of the 10 galaxies in our sample have sufficiently robust CO(3-2) data in the central regions to enable \mbh~estimations with our ML method: NGC~4388, NGC~5506, NGC~5643, NGC~6300, NGC~7314, NGC~7465, and NGC~7582. 
Before comparing our \mbh~estimations with those from the literature, we need to verify their reliability with the tools mentioned in section \ref{subsection:stats}.

\begin{table*}[h]
\caption{\label{table:stats}ML models statistics}
\hspace*{-0.1cm}
    \renewcommand{\arraystretch}{1.15}
    \begin{center}
    \begin{tabulary}{\linewidth}{|c|cc|cc|cc|}
    \hline
    \multirow{2}{*}{ID} & \multicolumn{2}{c|}{RMSE} & \multicolumn{2}{c|}{$R^2$} & \multicolumn{2}{c|}{$\overline{RE}$}\\
    \cline{2-7}
             & PVD    & Mom    & PVD   & Mom   & PVD    & Mom\\
    \hline
    NGC~4388 & 0.0341 & 0.0341 & 0.986 & 0.946 & 0.0020  & 0.0007 \\
    \hline
    NGC~5506 & 0.0335 & 0.0367 & 0.986 & 0.947 & 0.0060  & -0.0032 \\
    \hline
    NGC~5643 & 0.0194 & 0.0178 & 0.995 & 0.996 & -0.0019 & 0.0001 \\
    \hline
    NGC~6300 & 0.0243 & 0.0180 & 0.995 & 0.996   & -0.0002 & 0.0005 \\
    \hline
    NGC~7314 & 0.0312 & 0.0277 & 0.984 & 0.990 & -0.0029 & -0.0041 \\
    \hline
    NGC~7465 & 0.0246 & 0.0255 & 0.993 & 0.992 & 0.0010  & 0.0014 \\
    \hline
    NGC~7582 & 0.0263 & 0.0220 & 0.992 & 0.994 & -0.0002 & -0.0031 \\
    \hline
    \end{tabulary}
\tablefoot{RMSE, $R^2$ and mean RE of our models on the test sets. The RMSE ($>0$), $R^2$ (between 0 and 1) and the RE are defined in Section 4.5.} 
    \end{center}
\end{table*}

We report in Table~\ref{table:stats} the measure of the RMSE, the $R^2$, and the mean RE of our models on their test set for both the ones estimating the \mbh~ via the PVD and the ones using the first-moment maps. 
The analysis reveals that the diverse evaluations of the fitting quality exhibit a high degree of reliability and demonstrate consistency across the various indicators for both architectures. 
The RMSE values are at worst 0.0367 and the $R^2$ are all above 0.946, confirming that all the models are capable of making a reliable estimation of the \mbh~ on simulated data that were not present during the training process.

\begin{figure*}[h]
    \centering
     \includegraphics[width=1.0\linewidth,height=30cm,keepaspectratio]{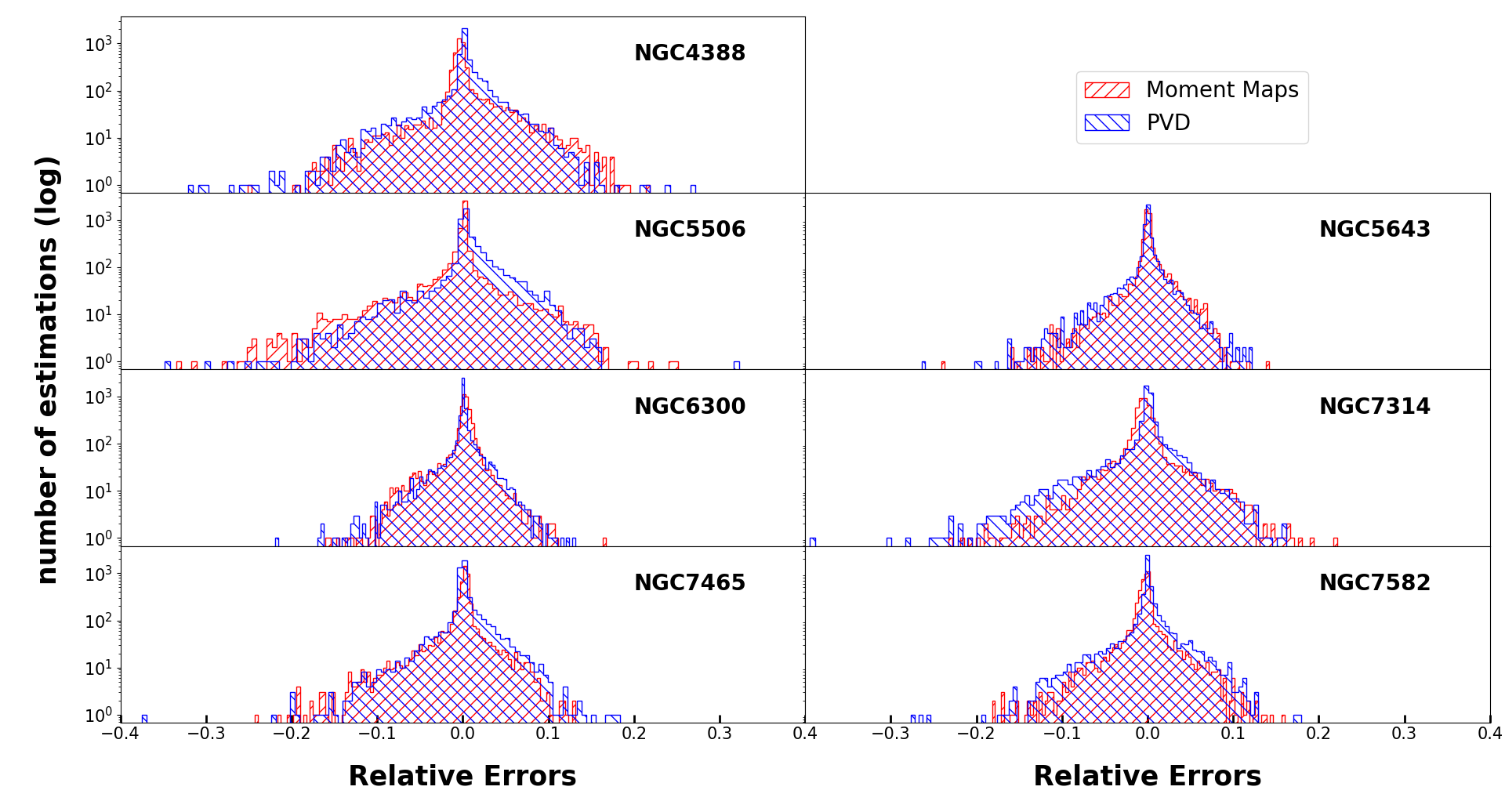}
    \caption{Histograms of the RE of our models when estimating the \mbh~ of the 5000 simulations from their respective test set. We represent in blue the models trained to estimate the \mbh~ with the PVD, and in red the models trained with the Moment maps.}
    \label{fig:mod_err}
\end{figure*}

We show in Fig~\ref{fig:mod_err}, the models RE distributions estimations on the 5000 simulations that composed their respective test sets. 
As shown in Table~\ref{table:stats}, the mean RE is $\sim$0 for the fourteen models, two for each galaxy, one with the PVD as input and one with the moment maps as input. 
The histograms are centered at the value of zero and prominently feature a pronounced peak at this center point.
The locations of the centers around 0 show that there is no systematic bias during the predictions and the relatively symmetrical distributions indicate that there is no constant over/underestimation of the \mbh.

The mock datasets used for training, validation, and testing are produced via a numerical simulation with 10 free parameters. We can use the RE to make sure that the confidence of our models \mbh~ predictions are not influenced by any other parameters than the \mbh. 
For each parameter, we plot the RE of the \mbh~ predictions as a function of the parameter values. We observe the same behavior for each galaxy so we present in the Appendix only the case of NGC~6300 (i.e. Fig~\ref{fig:err6300_details}). For every parameter, except for the \mbh, the distribution is uniform and centered around zero. This is strong evidence for considering that the models prediction does not depend on any other parameter other than \mbh.

To further investigate the behavior of our models, we represent the 2D histograms of the RE as a function of \mbh~ (i.e. Fig.\ref{fig:4388_hist2D} to \ref{fig:7582_hist2D}). 
All the distributions have the same shape.
From $log(M_{BH}/M_{\odot})$ of 4.0 to $\sim$4.5~dex, the distribution means increase linearly from $<0$ to $\sim$0 and their scatter stays constant. At $\sim$4.5~dex, the scatters get bigger and we see outliers that are underestimated. Then, the distribution scatters get thinner and the outliers number decreases as the \mbh~ predicted increases until $\sim$6.5~dex or $\sim$7~dex and stays constant afterward. 
Let us note the exception of the NGC~5643 model using the moment maps. Unlike the other models, its mean is already at 0 for $log(M_{BH}/M_{\odot})$=4~dex and stays constant with \mbh. Its scatter is also constant and uniform.   
As the \mbh~ gets lower, fewer and fewer pixels contain information about it. Thus, it becomes harder for the NN to extract meaningful information out of the data. This may explain why we observe the high scatter below $log(M_{BH}/M_{\odot})$ $\sim$6~dex.

\subsection{\texorpdfstring{\mbh}{} estimations results}\label{sec:GATOS res}

\begin{figure*}[h]
\centering
\includegraphics[width=1.0\linewidth,height=30cm,keepaspectratio]{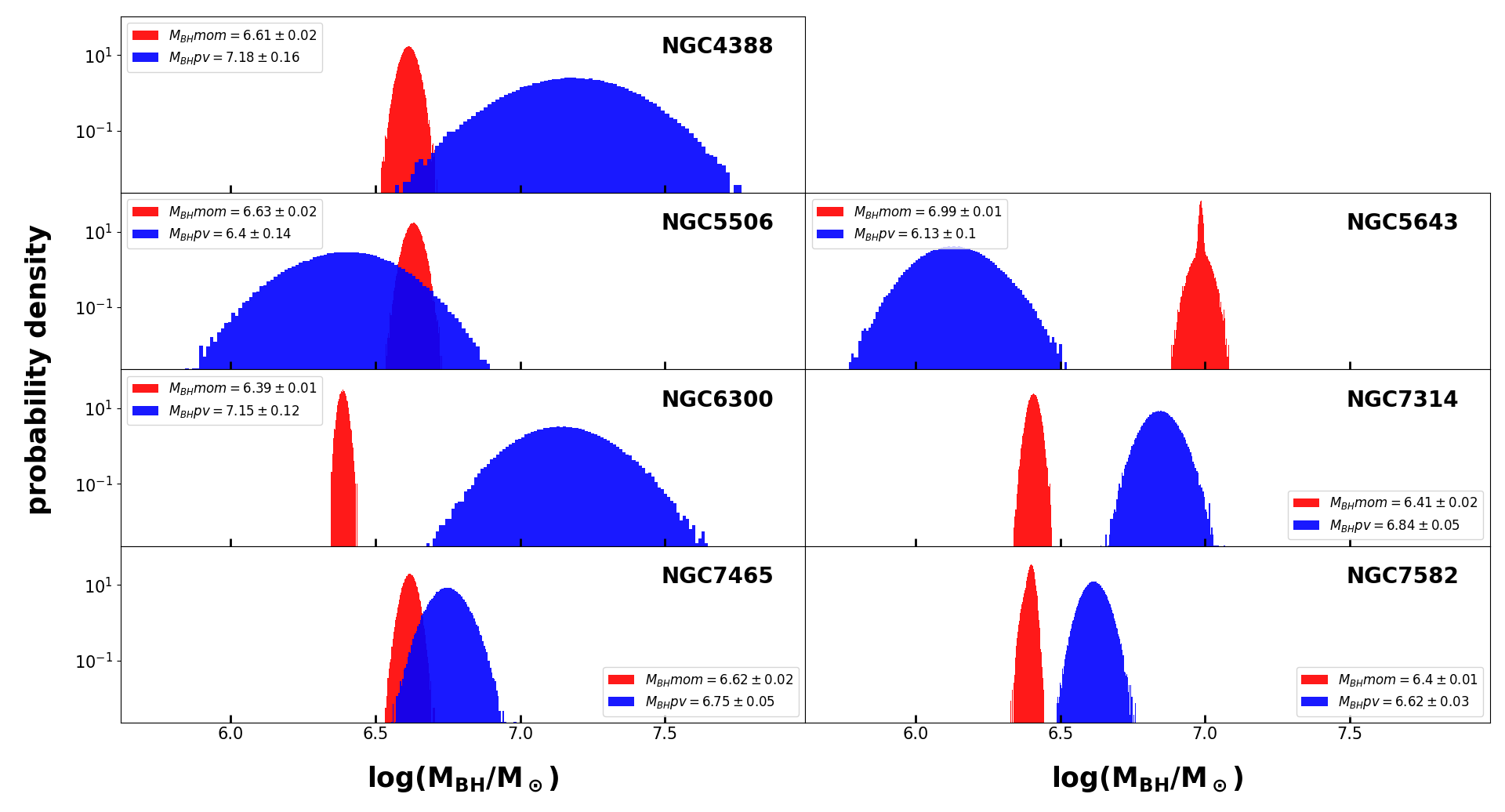}
\caption{Probability distributions of the \mbh~ predicted by our models for the seven galaxies in our sample. We have plotted the estimations made with the PVD as input in blue, and the estimations made with the first moment maps as inputs in red.}
\label{fig:est_mbh}
\end{figure*}

\begin{figure*}[h]
\centering
\includegraphics[width=1.0\linewidth,height=30cm,keepaspectratio]{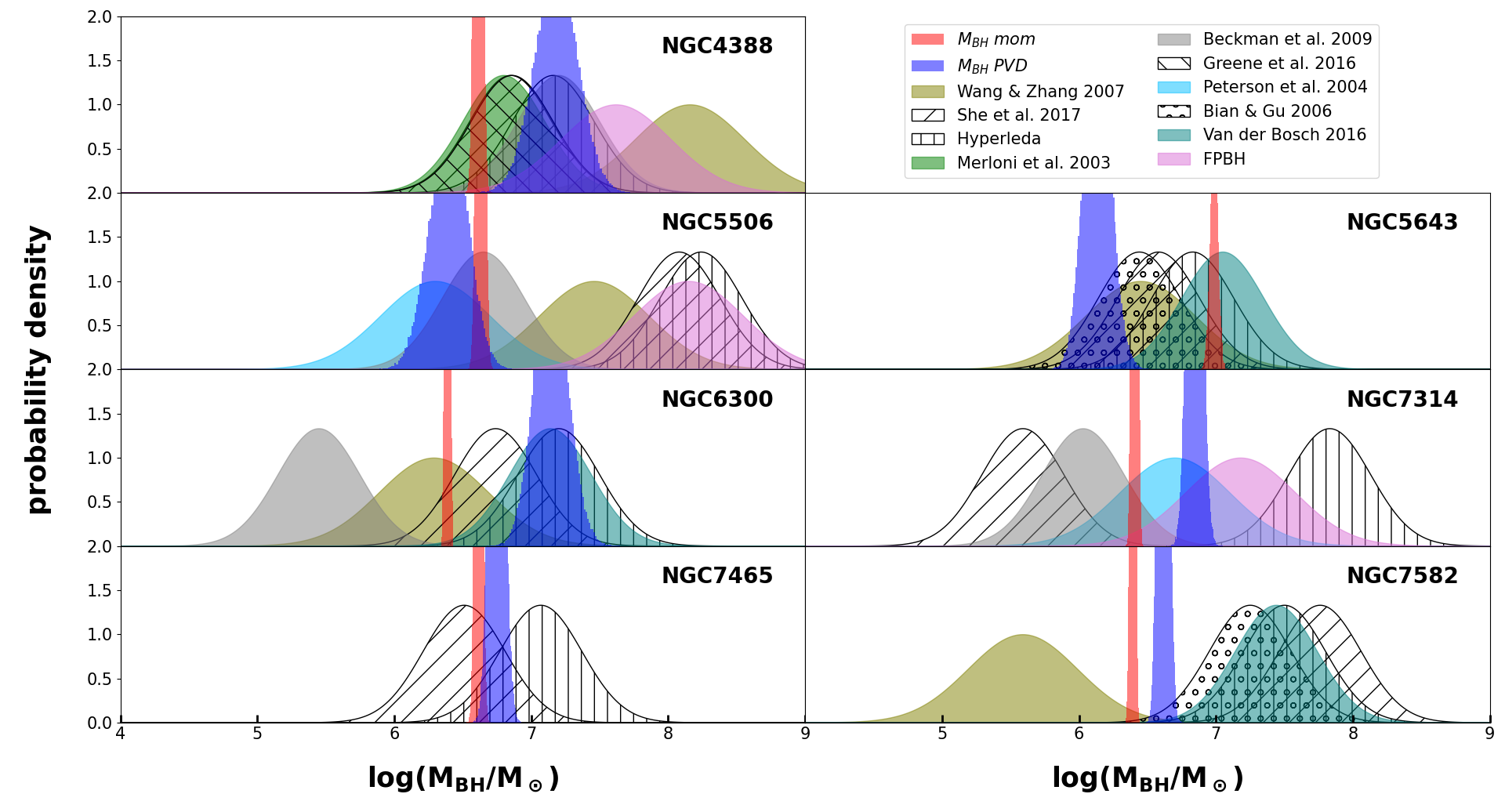}
\caption{Comparison of the probability distributions of the \mbh~ predicted by our models and the ones found in the literature. We represent in blue the results of the estimation made from the PVD and in red the estimation made from the moment maps. The other Gaussians represent the estimations found in the literature. The y-axis upper limit is fixed at 2 arbitrarily fixed at 2 for clarity purposes.}
\label{fig:comp_mbh}
\end{figure*}

\begin{table*}[h]
\caption{\label{table:lit_mbh}Summary of \mbh~ estimations}
	\vspace{-0cm}
	\hspace*{1cm}
	\centering
	\begin{adjustbox}{width=0.9\textwidth, left}
	\renewcommand{\arraystretch}{1.3}
	\scriptsize
	\begin{tabular}{ccccccccccc}
		\toprule
		ID      & log\mbh~ PVD    & log\mbh~ Mom    & M - $\sigma$         & Reverberation        & FPBH                  & Ref\\
		\midrule        
		NGC4388 & 7.18 $\pm$ 0.16 & 6.61 $\pm$ 0.02 & $7.17_{6.85}^{8.16}$ & ---                  & 7.62                  & 3,4,5,6,7,8   \\
		NGC4941 & ---             &      ---        & $6.83_{6.83}^{7.14}$ & 6.34                 & ---                   & 2,3,6,7       \\
		NGC5506 & 6.40 $\pm$ 0.14 & 6.63 $\pm$ 0.02 & $7.61_{6.65}^{8.24}$ & 6.30                 & 8.15                  & 1,3,4,6,7,8,9 \\
		NGC5643 & 6.13 $\pm$ 0.10 & 6.99 $\pm$ 0.02 & $6.73_{6.44}^{7.05}$ & $6.38_{6.30}^{6.45}$ & ---                   & 2,3,5,6,7     \\
		NGC6300 & 7.15 $\pm$ 0.12 & 6.39 $\pm$ 0.01 & $6.63_{5.45}^{7.14}$ & 6.29                 & ---                   & 3,4,5,6,7     \\
		NGC6814 & ---             &      ---        & $6.96_{6.42}^{7.32}$ & 7.28                 & 7.45                  & 1,4,5,7,9     \\
		NGC7213 & ---             &      ---        & $8.49_{8.06}^{8.82}$ & $7.35_{6.88}^{7.99}$ & ---                   & 1,3,7         \\
		NGC7314 & 6.84 $\pm$ 0.05 & 6.41 $\pm$ 0.02 & $6.48_{5.59}^{7.83}$ & 6.70                 & 7.18                  & 1,4,6,7,9     \\
		NGC7465 & 6.75 $\pm$ 0.05 & 6.62 $\pm$ 0.02 & $6.79_{6.51}^{7.07}$ & ---                  & ---                   & 6,7           \\
		NGC7582 & 6.62 $\pm$ 0.03 & 6.40 $\pm$ 0.01 & $7.56_{7.24}^{7.76}$ & 5.59                 & ---                   & 2,3,5,6,7     \\
		\bottomrule
		\end{tabular}
    \end{adjustbox}
    \tablefoot{Table summarizing the different \mbh~ estimations found in the literature and the one we made with ML. The \mbh~ is in $M_{\odot}$, and only the log of values are indicated. When multiple \mbh~ estimations using the same method were available, we report here their mean, the highest values in superscript, and the lowest in the subscript. 
    (1) Galaxies ID
    ; (2) Mean of the \mbh~ probability distribution obtained with ML and the PVD as input (the errors are at 1$\sigma$)
    ; (3) Same as the previous column but with the moment maps as input
    ; (3) Mean of the \mbh~ estimations with the M - $\sigma$ (intrinsic errors of 0.3 dex)
    ; (4) \mbh~ calculated with reverberation mapping (intrinsic errors of 0.4 dex)
    ; (5) FPBH masses (intrinsic errors of 0.4 dex)
    ; (6) \mbh~ estimations references 
    (1: \cite{peterson_central_2004},
    2: \cite{bian_eddington_2007},
    3: \cite{wang_millimeter_2007},
    4: \cite{beckmann_second_2009},
    5: \cite{van_den_bosch_unification_2016},
    6: \cite{she_chandra_2017},
    7: M-$\sigma$ estimation made with the Hyperleda data,
    8: FPBH mass estimated from the XMM-Newton and the FIRST data,
    9: FPBH mass estimated from the XMM-Newton and the NVSS data).} 
\end{table*}

The statistical indicators showed that the models succeeded in predicting precisely the \mbh~ on unseen simulated data when the $log(M_{BH}/M_{\odot})$ are above $\sim 6$~dex. 
So, if the numerical simulation can accurately emulate the observations, our models should be able to have the same behavior when estimating the \mbh~ of their corresponding observations. we show in Fig~\ref{fig:est_mbh} our models estimations on their respective observations. 
We made $10^{5}$ predictions per galaxy. 
The observed scatter on the \mbh~ predictions comes from the MC dropout. 
These scatters represent the models confidence in their predictions and do not have a predefined shape. 
The observed Gaussian shapes come from the fact that our models are confident in their estimations. 
Moreover, it's important to note that the models are trained on numerical simulations that do not encompass all the underlying physics observed in reality. Thus, the incertitude shown here cannot be associated with physical ones but only with the models confidence in their estimations. 

We can observe that the distributions are all Gaussian, except for the NGC~5643 model using the moment maps. 
This particular model seems to be the sum of two Gaussians with different dispersions but centered around the same value (6.99~dex). 
The width of the Gaussians given by the PVD models is broader compared to the moment maps models. Furthermore, the models scatter is constant for the moment map models, whereas it is galaxy-dependent for the PVD models. 
This can be explained if the PVD shows only a part of the gas dynamic. The effectiveness of the NN will rely on the extent to which the presented information is valuable for estimating the \mbh. On the contrary, the moment maps show the entirety of the gas dynamics, so helpful information is always represented. 
On one hand, the moment maps contain information primarily concentrated in the values of the central pixels, providing a localized representation of the \mbh-related information.
On the other hand, the PVD captures information through the velocity distribution as a function of position. Particularly, in the shape of this distribution in the vicinity of the center. This results in a more sparse distribution of information which makes it more difficult for the network to relate this information to the \mbh.
Another explanation is that we use different architectures for the PVD and the moment maps.

We can observe that, in Fig.\ref{fig:comp_mbh}, for four out of the seven galaxies, the estimations of the PVD and the moment maps do not overlap. This can be attributed to the same hypotheses as to why the PVD distributions are broader than the moment maps.

We compare our \mbh~ estimations with the ones previously made using classical methods in Table~\ref{table:lit_mbh} and Fig~\ref{fig:comp_mbh}. For the seven galaxies in our sample, the \mbh~ predicted by our models is coherent with previous estimations. There exists some overlap with at least one previous estimation, except for NGC~7314 and NGC~7582. For them, our estimations fall in between the one by \cite{wang_unified_2007} and the three others made by \cite{bian_eddington_2007,van_den_bosch_unification_2016,she_chandra_2017}.

\subsection{Comparison with classical method}

In this study, we used 25,000 moment maps of NGC~7582 previously simulated datasets, originally generated for ML methods, to explore traditional non-ML approaches for estimating \mbh, specifically linear interpolation and chi-squared fitting. 
These non-ML methods provided a useful comparison to ML models, as they rely on conventional estimation techniques rather than data-driven learning. Linear interpolation estimates unknown values by assuming a linear relationship between nearby data points, while chi-squared fitting minimizes the discrepancies between observed and predicted data. Both methods were applied to these simulations to evaluate their performance in estimating \mbh.

\begin{figure*}[h]
\centering
\includegraphics[width=0.7\linewidth,height=30cm,keepaspectratio]{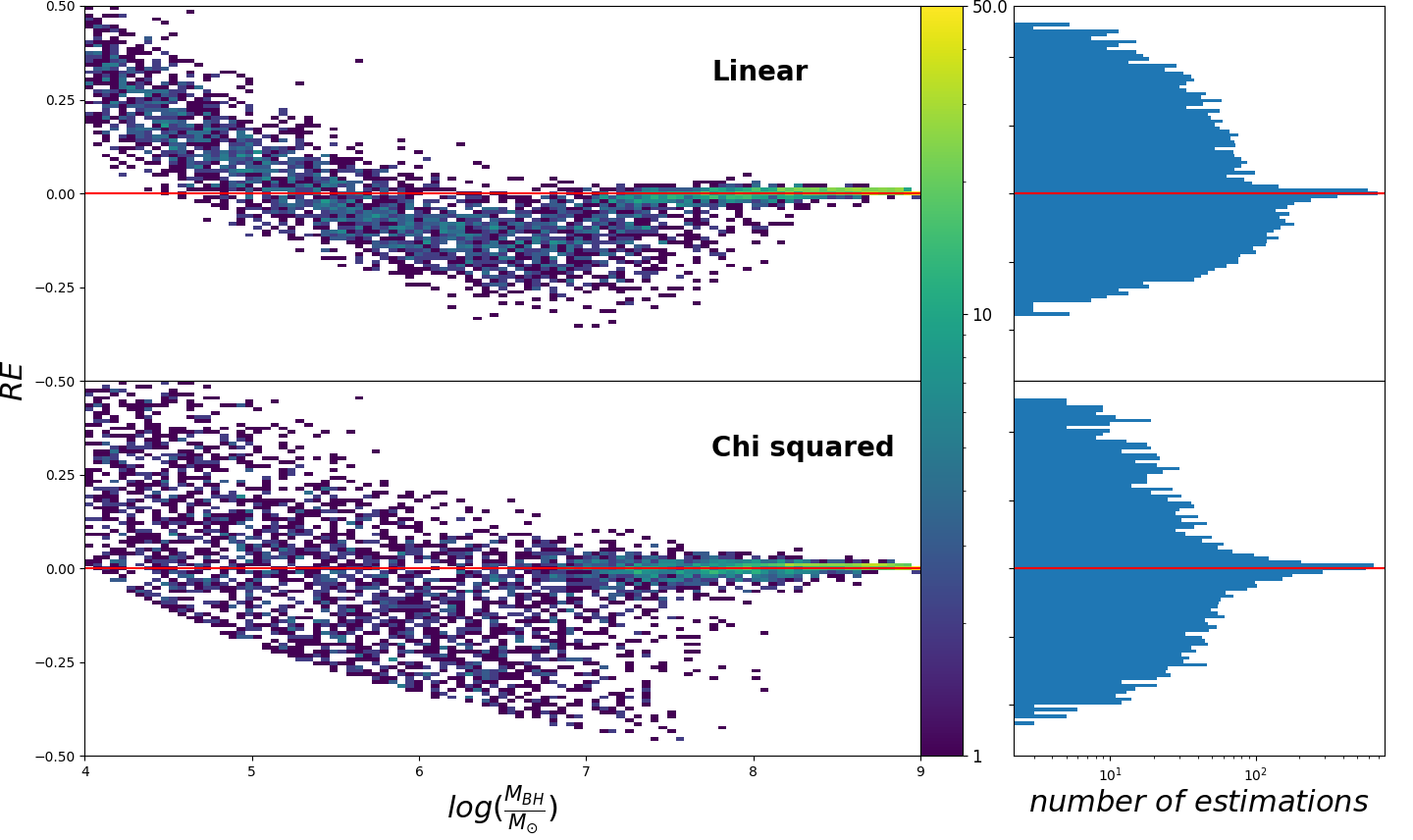}
\caption{Two-dimensional and one-dimensional histograms of the RE from linear interpolation and chi squared fitting of \mbh~ on the test set for NGC~7582. 
 On the left, we plotted the RE when estimating the \mbh~ of the test set as 2D histograms.
 On the right, we plotted the same RE but as a 1D histogram. 
 On top, we plotted the RE from the linear interpolation and at the bottom, we plotted the RE from the chi squared fitting.
 The y-axis represents the RE values. The x-axis of the 2D histograms is in $log(\frac{M_{\mathrm{BH}}}{M_{\odot}})$ and the 1D histograms axis is in number of estimations.}
\label{fig:linear_err}
\end{figure*}

Linear interpolation works by assuming that the changes between adjacent data points are linear, allowing unknown values within the dataset to be estimated. In our simulations, the data was split into 20,000 training points and 5,000 test points. However, due to the high dimensionality of the data, memory limitations arose when attempting to interpolate across the entire dataset. To overcome this, we adopted a localized interpolation strategy. Instead of interpolating over the full dataset, we identified the K nearest neighbors in multidimensional space—those points closest to the test data—and performed linear interpolation on this subset. This localized approach reduced both memory usage and computation time.
Despite its efficiency, the linear interpolation method was less accurate than the ML models. As shown in Fig.~\ref{fig:linear_err}, the histogram of RE from linear interpolation indicates a peak near zero, but the errors were generally higher than those observed with ML estimations. Moreover, the error distribution was asymmetric, implying that the linear assumption in high-dimensional space may not fully capture the data's complexity.
The 2D histograms of RE provide additional insight into the limitations of linear interpolation (Fig.~\ref{fig:7582_hist2D}). Unlike the ML model, which consistently estimated \mbh~ starting from around 5.5 dex, the linear interpolation method only showed reliable performance at the upper end of the \mbh~ distribution, approximately from 8 dex onwards. This suggests that linear interpolation struggles to produce accurate estimates across the entire \mbh~ range, particularly in the lower regions where relationships between variables are likely more complex and non-linear. As a result, while linear interpolation may offer computational efficiency, its ability to generalize across different regions of the \mbh~ distribution is limited, making it less reliable than ML approaches for comprehensive mass estimation.

Chi-squared fitting in 2D works by comparing observed data points to a theoretical model and minimizing the differences between them. For each pair of observed and expected values, the squared difference is calculated and normalized by the variance of the observed data. These individual terms are then summed to produce the chi-squared statistic, which quantifies the overall goodness of fit. The method iterates through possible model parameters, adjusting them to reduce the chi-squared value.
Figure \ref{fig:linear_err} presents 2D histograms illustrating the distribution of chi-squared results, which exhibit a similar shape to those obtained using linear interpolation. Both methods display a comparable constant scatter, not centered around zero, indicating a shared unreliability in estimating \mbh~ below approximately 8 dex. Notably, the scatter in the lower mass range is more pronounced in the chi-squared 2D histograms. However, the 1D histograms reveal that the RE distribution for the chi-squared method is more symmetric, suggesting improved consistency in error distribution when compared to linear interpolation.

In conclusion, both the linear interpolation and chi-squared fitting methods exhibit significant challenges in reliably estimating \mbh~ within the 4 to 8 dex range. In contrast, our ML approach demonstrates improved performance, effectively mitigating this limitation and confining the estimation inaccuracies to a narrower range of 4 to 6 dex. This highlights the enhanced robustness and precision of the ML method, particularly for mass estimates in the lower dex intervals.



\section{Individual galaxies}

The three moments of the CO(3-2) cubes from ALMA observations (top panel) are compared with the best fit of our simulations (bottom panel) on the left sides of Figs.~\ref{fig:NGC4388_mom} to \ref{fig:NGC7582_mom}. On the right side, the observed PVD is overlaid with the  contours of the second model, optimized for this. The models are fitted with the same methodology as in \cite{combes_alma_2019}. 

\subsection{NGC~4388}

The galaxy is almost edge-on ($IA$ of 79$^{\circ}$), and its nuclear molecular disk reveals a ring of $\sim$ 100 pc radius, which corresponds to the inner Linblad resonance of the bar. The ring is not totally empty and shows clumps that enter the SoI, allowing a \mbh~ estimation. Some molecular gas is expelled along the minor axis and might be dragged by the ionized wind and radio jet \citep{Falcke1998}.

Figure \ref{fig:NGC4388_mom} shows the model best fit found for NGC~4388.
It is clear in this figure that the velocity field of the galaxy is highly asymmetric, and that our symmetric gravitational potential model cannot reproduce the observed first-moment map. While the gas with negative velocity distribution is well reproduced, the positive part is not.  On the PVD panel, the model fits better than the symmetrized observation. The observed second moment shows a sudden drop of velocity dispersion at P$\sim$200pc that is not present in the model. Moreover, the model has a large velocity gradient at the center contrary to the observations. This comes from the lack of gas in the SoI, so it is difficult to precisely measure the SMBH mass in this galaxy. It can be noted that the lowest red contour in the PVD is less reliable, related to a low level of emission.

\subsection{NGC~5506}
 
NGC~5506 is also almost edge-on ($IA$ of 80$^{\circ}$), and possesses a nuclear molecular ring of radius $\sim$ 50 pc, corresponding to the ILR of its strong bar. There is however dense molecular gas near the AGN, and HCO$^+$ has been detected in the center \citep{garcia-burillo_galaxy_2021}. The estimation of \mbh~ is therefore possible.

 Figure \ref{fig:NGC5506_mom}, reveals in the center of the observed first-moment map that the gas velocity distribution has an S-like shape. This is probably due to the presence of a bar. Except for these details, the model can well reproduce the observed gas velocity distribution. However, it cannot do as well for the velocity dispersion distribution. The model gives a more or less constant velocity dispersion whereas two clumps with high-velocity dispersion are observed near the center and a low-velocity dispersion elsewhere.
 This effect is also seen in the PVD and could be also due to the bar. The low gas density in the center makes the SMBH mass estimation uncertain. This is reflected in the low-emission red contour of the model, which proposes in general a higher \mbh~ than the moment-based model.

\subsection{NGC~5643}

This is an almost face-on barred spiral, with an $IA$ of 29$^{\circ}$. The strong stellar bar is almost E-W, with PA= 85$^{\circ}$. At $\sim$100~pc scale, NGC~5643 CO(3-2) moment zero map shows the presence of a nuclear bar, prolonged by a spiral structure, but the CO bar is almost N-S, with PA= 5$^{\circ}$, cf Fig~\ref{fig:NGC5643_mom}.
The very center reveals a tilted molecular torus, of diameter 26~pc, oriented almost N-S, perpendicular to the ionized gas outflow and radio jet \citep{Alonso-Herrero2018,Garcia-Bernete2021}. Dense gas traced by HCO$^+$ has been detected in the very center \citep{garcia-burillo_galaxy_2021}. 

The NGC~5643 model reproduces well the shift in the velocity orientation towards the nucleus (Fig.\ref{fig:NGC5643_mom}). However, the second-moment map reveals a high-velocity dispersion in the nucleus, which is horizontal in the model while almost vertical in the observation.
In the PVD, the model does not distribute the gas clumps at the same positions as in the observed diagram; however,  the velocity gradient in the nucleus is reproduced, and the SMBH mass estimation appears robust. Note that the moment-based model leads to a higher \mbh~  in this galaxy, which is unusual, maybe due to the strong misalignment.

\subsection{NGC~6300}

This is a strongly barred galaxy, moderately inclined (57$^{\circ}$). At $\sim$100~pc scale, the CO(3-2) emission shows a detached nuclear disk, with a central concentration, allowing a determination of \mbh, cf Fig~\ref{fig:NGC6300_mom}. This strong concentration corresponds to a molecular torus of 25~pc radius, also detected in HCO$^+$. The torus oriented at $PA=85^{\circ}$, is perpendicular to the ionized gas outflow, detected by \cite{Julia2016}.

For the NGC~6300 fit (Fig.\ref{fig:NGC6300_mom}), the first-moment map is well reproduced by the model, except for the wiggle at the center of the disk, likely due to the bar. The modeled second-moment map underestimates the velocity dispersion at the center but overestimates it near the edges. Furthermore, the central region with high-velocity dispersion is horizontal in the observation while it is almost vertical in the model, a likely consequence of the bar. The PVD panel reveals a reasonable fit, with a sufficient gas sampling of the SoI.

\subsection{NGC~7314}

This barred spiral galaxy looks relatively inclined (70$^{\circ}$) on the sky, but is kinematically less (55$^{\circ}$) since its morphological shape is biased by the bar. At $\sim$100~pc scale, the molecular disk reveals a central depletion, surrounded by two peaks, suggesting the existence of a ring of radius 20~pc, that could be an ILR resonant ring. Its PVD shows very little gas in the SoI but it is enough to estimate an upper limit of 6.2 dex on the log (\mbh/M$_\odot$). In this galaxy, the molecular nuclear disk does not appear strongly misaligned with the large-scale one. It is also oriented along the radio source \citep{Thean2000}, which might be dominated by star formation.

The NGC~7314 fit (Fig.\ref{fig:NGC7314_mom}) overestimates the velocity gradient in the outer parts of the disk. The model also overestimates the central velocity dispersion along the minor axis, while the dispersion is more spatially distributed in the observation. The PVD reveals a high central velocity dispersion, leading to a high estimation of the SMBH mass. This is however provided by the lowest emission red contour.

\subsection{NGC~7465}

This barred early-type spiral is moderately inclined (60$^{\circ}$) on the sky and its morphological major axis of $PA= 160^{\circ}$ at large-scale is roughly perpendicular to the molecular disk at $\sim$100~pc scale, with $PA= 45^{\circ}$. The circum-nuclear disk in the regular rotation is dense, of 70~pc radius, and prolonged by a sparse spiral structure. There is enough dense gas towards the center to estimate \mbh~ through the ML method. The inferred log($M_{\mathrm{BH}}/M_\odot$) is 5.9 $\pm$ 0.8 dex.

The model for NGC~7465 (Fig.\ref{fig:NGC7465_mom}) reproduces very well the observed first-moment map. The velocity dispersion is however gathered along the minor axis, while it is more spatially distributed in the observed second moment. The low-velocity gradient in the SoI region leads to a low mass estimation for the SMBH.

\subsection{NGC~7582}

NGC~7582 is a barred early-type spiral, highly inclined (68$^{\circ}$), with a $\sim$100~pc scale molecular disk aligned with the large-scale galaxy disk, with $PA = 156^{\circ}$. The molecular gas is condensed in a 200~pc radius ring, which corresponds to an ILR ring. However, there is enough CO emission right into the very center, making it possible to estimate \mbh. There is even a concentration of CO emission in a small disk of radius 30~pc, corresponding to a molecular torus. An outflow of ionized gas is observed in the perpendicular direction \citep{Davies2016}.

The NGC~7582 model fits the observed velocity gradient in the disk relatively well, although its morphology is too regular (Fig.\ref{fig:NGC7582_mom}). The observed first-moment map shows that the iso-velocity morphology gets thinner towards the nucleus, with a large central velocity gradient, which might be due to the bar. The velocity dispersion reflects also this morphology and is not reproduced in the model. The PVD reveals a low-velocity gradient in the SoI and a low SMBH mass estimation.

\section{Discussion}

To provide more robust estimations of \mbh, there are three ways of improvement: the input data, the numerical simulations, and the ML method itself. 
First of all, the model precision is limited by the observation's spatial resolution. 
Figure \ref{fig:4388_hist2D} demonstrates that NN face challenges in providing accurate estimations when the $log(M_{BH}/M_{\odot})$ is below approximately $6$ dex. This issue arises because, as \mbh~ decreases, fewer pixels in the data carry information about it. Consequently, the NN struggles to extract meaningful insights, leading to increased variability in estimations for smaller \mbh~ values. 
Therefore, observations with higher resolutions provide more pixels with relevant \mbh~ information, enhancing the NN precision and reducing the sensitivity to small \mbh~ variations. 
The role of the SoI in the context of NN remains somewhat unclear. While the goal is to increase the number of informative pixels about the BH, these pixels don't need to be situated where the BH dominantly influences gas dynamics. Rather, these pixels need only exhibit a noticeable effect on the NN, even if it's relatively small.

Another limiting factor comes from the physics included in the numerical simulations used to make the datasets. We use numerical simulations to mimic real observations but we simulated an axisymmetric and homogeneous thin gas disk, with no clumpy structure, nor spiral or bar structures. 
Furthermore, we have adopted in our models the same $IA$ and $PA$ for the gas and stellar disks, while some misalignments of the nuclear molecular disks have been observed with ALMA \citep{combes_alma_2019}. 
Adding these misalignments and more complex physics, like the presence of a bar, in the simulations would add free parameters to influence directly the \mbh~ estimations. This discrepancy in realism can result in the network exhibiting excessive confidence in its predictions, while there is a strong likelihood that the uncertainty in the actual data is being underestimated.

Additionally, as previously elaborated in Section \ref{sec:GATOS res}, the probability distributions illustrated in Fig.\ref{fig:est_mbh} lack physical significance. This discrepancy arises from the inherent differences in the underlying physics between the numerical simulations and the actual observational data. However, by applying a reverse perspective, as the physical fidelity of the simulations approaches that of the observations, the probability distributions of \mbh~ become more representative of the actual physical uncertainties. Therefore, enhancing the realism of the numerical simulations serves a dual purpose, it not only improves the accuracy of error estimates but also confers them with a deeper and more grounded physical significance.

The ML method in itself can also be improved. Here, we chose to work with one galaxy at a time which means creating a different training sample and training a different model for each galaxy. This approach has the advantage of simplifying the training set creation and the learning process but it is also very inefficient time-wise and this approach is effective solely due to the limited number of galaxies available in our dataset. It takes $\sim$ 8 hours to create a dataset and $\sim$ 4 hours to train one NN.

To enhance the efficiency and applicability of our method, it is essential to generalize it to work across multiple galaxies rather than being constrained to individual ones. By creating a training sample that includes moment maps from multiple different galaxies, we can develop a NN capable of generalizing and working for any galaxy. Generalization requires a robust training dataset that includes moment maps and PV diagrams from a wide variety of galaxies. This diversity is crucial for the NN to learn the different galactic structures, dynamics, and environments it might encounter.

Moreover, when working with multiple galaxies simultaneously, we expect the network to be able to mutualize or share learned features across different galaxies. This allows the NN to extract common patterns and relationships between galactic properties, enhancing its ability to generalize. Consequently, while the architecture needed for such a task might be more complex due to the need to handle diverse galactic environments, it does not necessarily have to be much larger in terms of parameters. The mutualization of features could lead to a more efficient learning process, allowing the network to converge without significantly increasing its size. Therefore, the inference time should not see a substantial increase, ensuring that the method remains computationally feasible even as it scales to a broader range of galaxies.

However, our current limited sample size allows us to center the moment maps and PV diagrams around the BH, a process that becomes challenging with a larger number of galaxies. To address this, incorporating a BH localization step within the network can further improve performance. This step involves identifying the position of the BH in the moment maps and PV diagrams before estimating its mass, enabling the model to handle cases where the BH is not centrally located.

To effectively learn and predict additional physical parameters and/or accurately detect the position of BH, it is necessary to make modifications to the NN architecture to address these challenges. 
Moreover, a larger and more diverse training sample will be required. When aiming to predict multiple parameters, it is essential to ensure that the training sample is sufficiently extensive to comprehensively map the entire feature space. 
Furthermore, to successfully learn how to localize the BH in the input data, the positional information of the BH must be known for each example. However, the specific size and diversity requirements of the sample are empirical and contingent upon the particularities of the problem at hand. At the very least, several hundred distinct galaxy molecular gas observations are anticipated to be necessary.

Then we can choose to train our NN on the observations or simulations made from the observations. 
Training the NN on the observations has the advantage that we do not have problems with the realism of the physics used in the numerical simulations.  However, it requires to have \mbh~ estimation for all the observed galaxies. 
In addition, it's important to note that a single galaxy may yield multiple \mbh~ estimations derived from various methods, often resulting in discrepancies between these estimates. This situation highlights the necessity for methodological consistency in estimating \mbh~ across all galaxies in the sample to avoid potential confusion for the NN. 
However, this uniformity in estimation methods also implies that the NN is not learning to estimate \mbh~ per se but is instead learning to estimate \mbh~ based on the predetermined method. Consequently, if the selected method contains inherent flaws or inaccuracies, the NN is likely to incorporate and replicate these shortcomings in its estimations. 
Working with numerical simulations does have its disadvantages, particularly concerning the intricacies of simulating the physics, as elaborated earlier. However, it also presents the advantage of reducing the volume of observations necessary for the method generalization, as it allows for the creation of multiple simulations per galaxy.

Nevertheless, obtaining the exact position of the BH may not be a prerequisite; instead, it may suffice to ensure that the BH is contained within the image. This approach offers the advantage of simplifying the acquisition of the required sample. However, it is important to note that training the model using this method may be harder.

\section{Conclusion}\label{sec:GATOS conclu}

The ALMA high sensitivity and spatial resolution allow us to resolve the CO(3-2) molecular disks at the center of nearby galaxies. In the case of the GATOS core sample used in this work, the distances and ALMA angular resolutions allow for resolution of typical physical scales of 10 pc, which probe their SoI.
Since the nuclear disks entered the sphere of influence of the BH, we were able to make \mbh~ estimations. In this article, we provide a proof of concept for a novel approach to estimate the mass of SMBH using ALMA observations of the circum-nuclear disk. Our method employs supervised ML techniques on data obtained from numerical simulations.

Gas dynamics can be studied by looking at the PVD and/or the moments map calculated from a spectral data cube. we developed two artificial NN architectures. The first one takes the PVD as input, the other one takes the first moment maps, and both of them give an estimation of the \mbh~ based on the input. 
To test the reliability of our models, we used three statistical indicators: the RMSE, the $R^2$, and the $\overline{RE}$. 
These statistical indicators confirmed that our fourteen models can make reliable \mbh~ estimations on new data. 
we looked at our models RE distribution when predicting the \mbh~ of the test sample. we saw that the distributions were symmetrical and centered around 0, indicating no constant over/underestimation. 
We also studied the models RE of their \mbh~ estimations made on the test set as a function of the simulation parameters values. The resulting distributions are all uniform and center around zero except for the \mbh. 
This is a strong indication that the models \mbh~ predictions only depend on the \mbh. 
We studied more in-depth the behavior of the \mbh~ predictions by looking at the predictions RE versus the \mbh.

we managed to make a \mbh~ estimation with both architectures for 7 out of the 10 galaxies in the GATOS core sample: NGC~4388, NGC~5506, NGC~5643, NGC~6300, NGC~7314, NGC~7465, and NGC~7582.  
we made $10^{5}$ predictions per galaxy using MC dropout. The resulting \mbh~ estimations are Gaussian-like probability distributions. The scatter of these distributions represents the models confidence in their predictions. we observe that the PVD models give larger scatters from 0.03~dex to 0.16~dex depending on the galaxy. On the other hand, the moment maps models have a constant scatter of $\sim$0.02~dex for all galaxies. The narrow error bars observed in our results are likely indicative of the network's overconfidence, which can be attributed to its training on numerical simulations that encompass simplified physics compared to real-world conditions. 
Nonetheless, all the estimations made are consistent with the previous studies except for NGC~7314 and NGC~7582 where our estimate is between the previous ones.

In this study, we showed that even with training sets simulated with simple physics, with a fixed $IA$ and $PA$, and with simple NN architectures, our approach can produce results coherent with the literature. 
This work represents the first step toward an automatized method for estimating \mbh.

In this study, we demonstrated that even with training sets simulated using simple physics, fixed $IA$ and $PA$, and basic NN architectures, our approach can produce results consistent with those in the literature. However, a key limitation of this work is that we trained and applied the NNs on one galaxy at a time. As a result, we had to generate numerical simulations and train a separate model for each individual galaxy. This approach limits scalability and the ability to generalize across different galactic environments. While our method is not necessarily superior to classical methods, it serves as a proof of concept and an initial step towards developing ML techniques for \mbh~ estimations that can better scale with the increasing number of astronomical observations.

\begin{acknowledgements}

We would like to thank the referee for very useful comments, that helped us improve and clarify the paper.
RP acknowledges financial support from the SNSF under the Weave/Lead Agency RadioClusters grant (214815).
AAH and MVM acknowledge support from PID2021-124665NB-I00 by the Spanish Ministry of Science and Innovation/State Agency of Research MCIN/AEI/ 10.13039/501100011033 and by ‘ERDF A way of making Europe.
SGB acknowledges support from the research project PID2019-106027GA-C44 of the Spanish Ministerio de Ciencia e Innovaci\'on. 
Training computation and inferences have been performed on the MINERVA group from the Observatoire de Paris GPUs server.
AJB has received funding from the European Research Council (ERC) under the European Union’s Horizon 2020 Advanced Grant 789056 “First Galaxies”.
Based on observations made with the NASA/ESA Hubble Space Telescope, and obtained from the Hubble Legacy Archive, which is a collaboration between the Space Telescope Science Institute (STScI/NASA), the Space Telescope European Coordinating Facility (ST-ECF/ESA), and the Canadian Astronomy Data Centre (CADC/NRC/CSA). MPS acknowledges funding support from the Ram\'on y Cajal program of the Spanish Ministerio de Ciencia e Innovaci\'on (RYC2021-033094-I).
C.R. acknowledges support from the Fondecyt Regular grant 1230345 and ANID BASAL project FB210003.
OGM acknowledges support from UNAM PAPIIT project IN109123 and CONACyT “Frontera de la Ciencia” project CF-2023-G-100.
BGL acknowledges support from grants PID2019-107010GB-100 and the Severo Ochoa CEX2019-000920-S funded by MICINN-AEI/10.13039/501100011033.
CRA acknowledges financial support from the European Union's Horizon 2020 research and innovation program under Marie Sk\l odowska-Curie grant agreement No 860744 (BiD4BESt), from the State Research Agency (AEI-MCINN) and from the Spanish MCINN under grants ``Feeding and feedback in active galaxies", with reference PID2019-106027GB-C42, the project ``Quantifying the impact of quasar feedback on galaxy evolution'', with reference EUR2020-112266, funded by MICIN/AEI/10.13039/501100011033 and the European Union NextGenerationEU/PRTR.

\end{acknowledgements}

\bibliographystyle{aa}
\bibliography{bibli}

\FloatBarrier
 \begin{appendix}
 \onecolumn
 \section{Moment maps and position-velocity diagrams} \label{app:contmaps}

\begin{figure*}[h]
\centering
\subfloat{}{\includegraphics[width=0.49\textwidth]{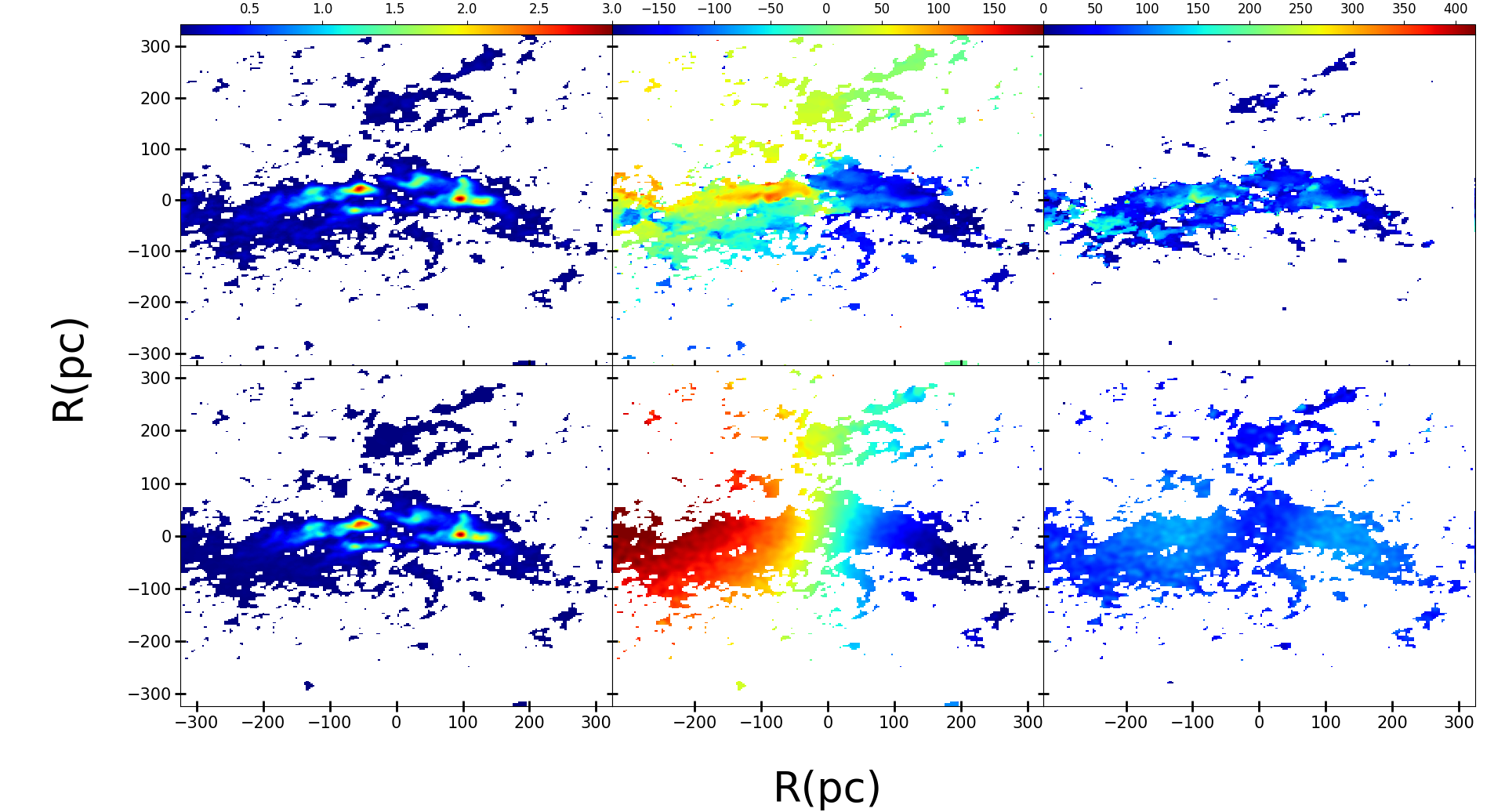}} \hfill%
\subfloat{}{\includegraphics[width=0.49\textwidth]{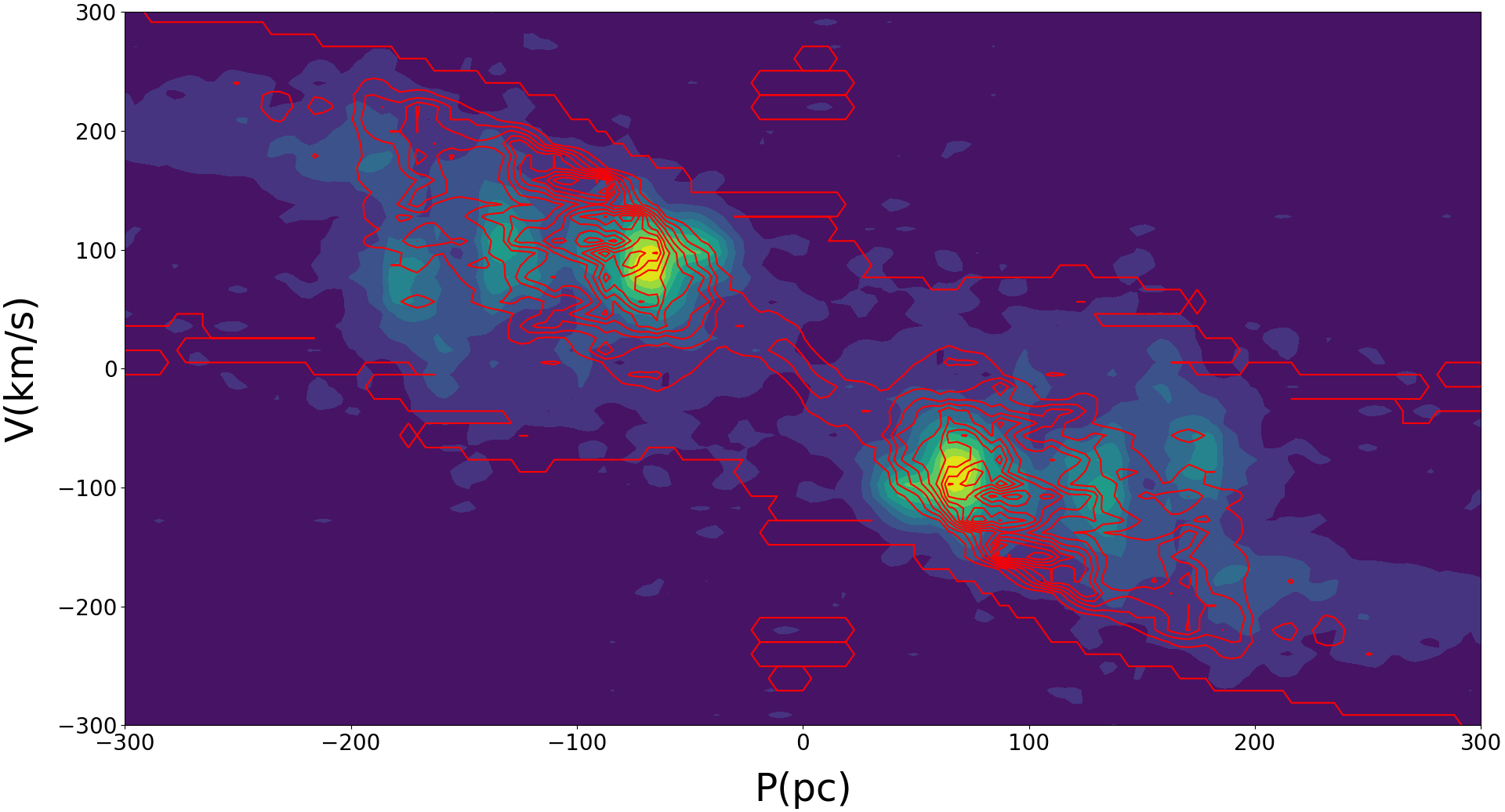}} 
\caption{Left we present the first three moments maps of NGC~4388, obtained from the best-fit model, giving the adopted \mbh, compared with the NGC~4388 CO(3-2) molecular gas deduced from the \cite{garcia-burillo_galaxy_2021} ALMA observations. From left to right, the integrated intensity, the velocity field, and the velocity dispersion; the top is the observation, the bottom is the best-fit model. On the right, we present the observed PVD, with the best-fit model overlaid in red contours. The parameters used for the fit are presented in Table \ref{table:fits_params}.}
\label{fig:NGC4388_mom}
\end{figure*}

\begin{figure*}[h]
\centering
\subfloat{}{\includegraphics[width=0.49\textwidth]{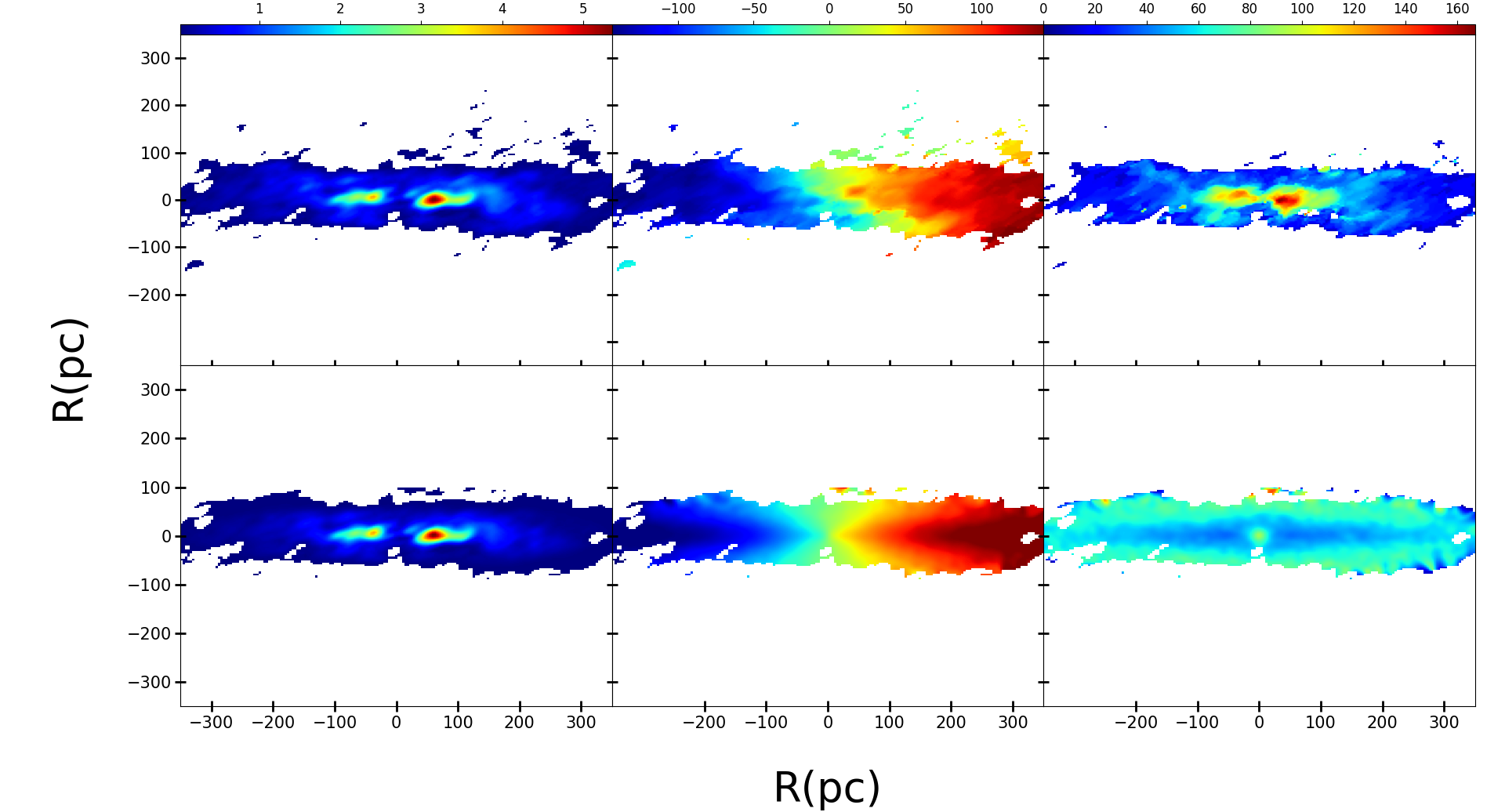}} \hfill%
\subfloat{}{\includegraphics[width=0.49\textwidth]{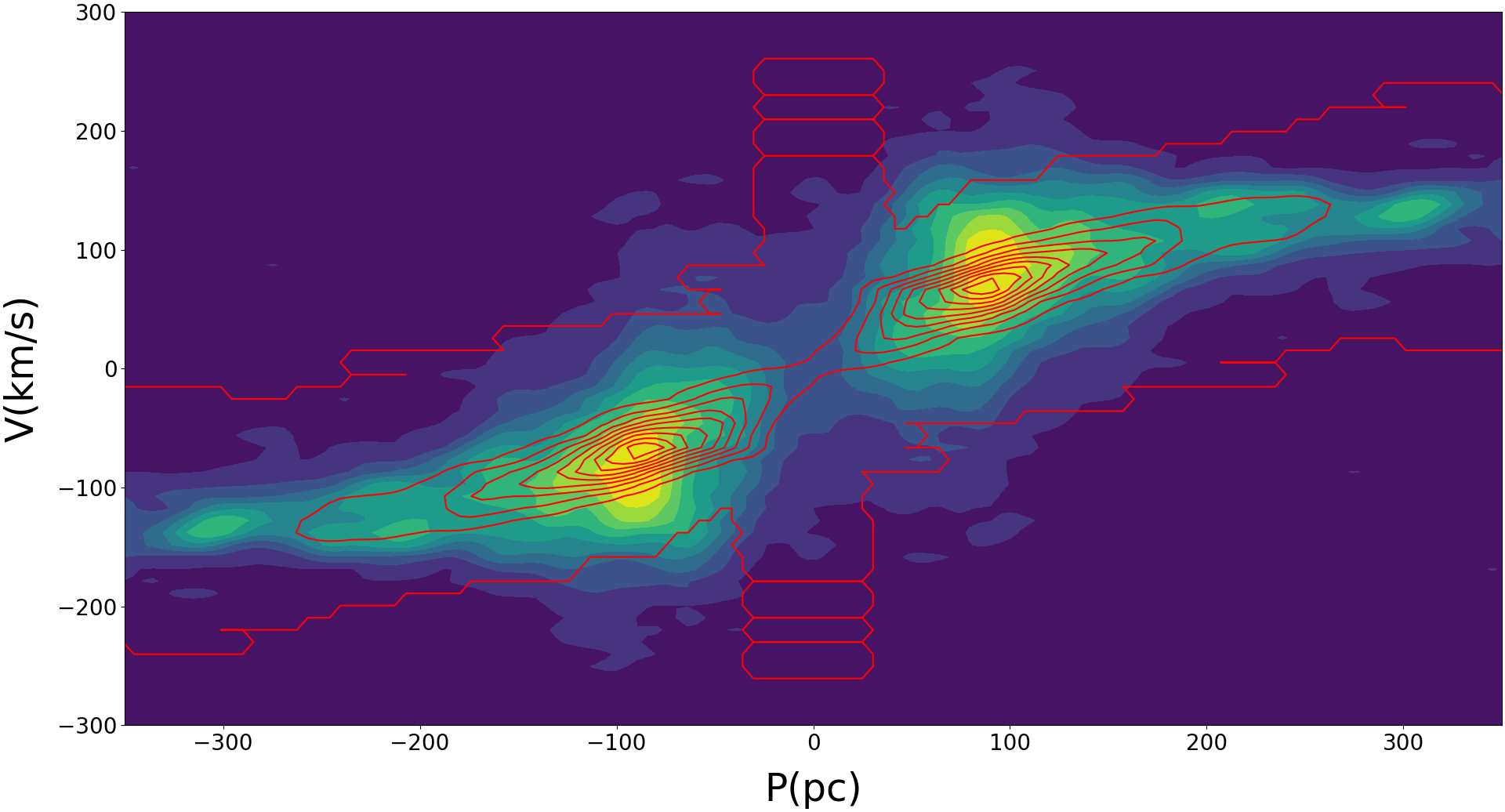}} 
\caption{Same as \ref{fig:NGC4388_mom} for NGC~5506.}
\label{fig:NGC5506_mom}
\end{figure*}

\begin{figure*}[h]
\centering
\subfloat{}{\includegraphics[width=0.49\textwidth]{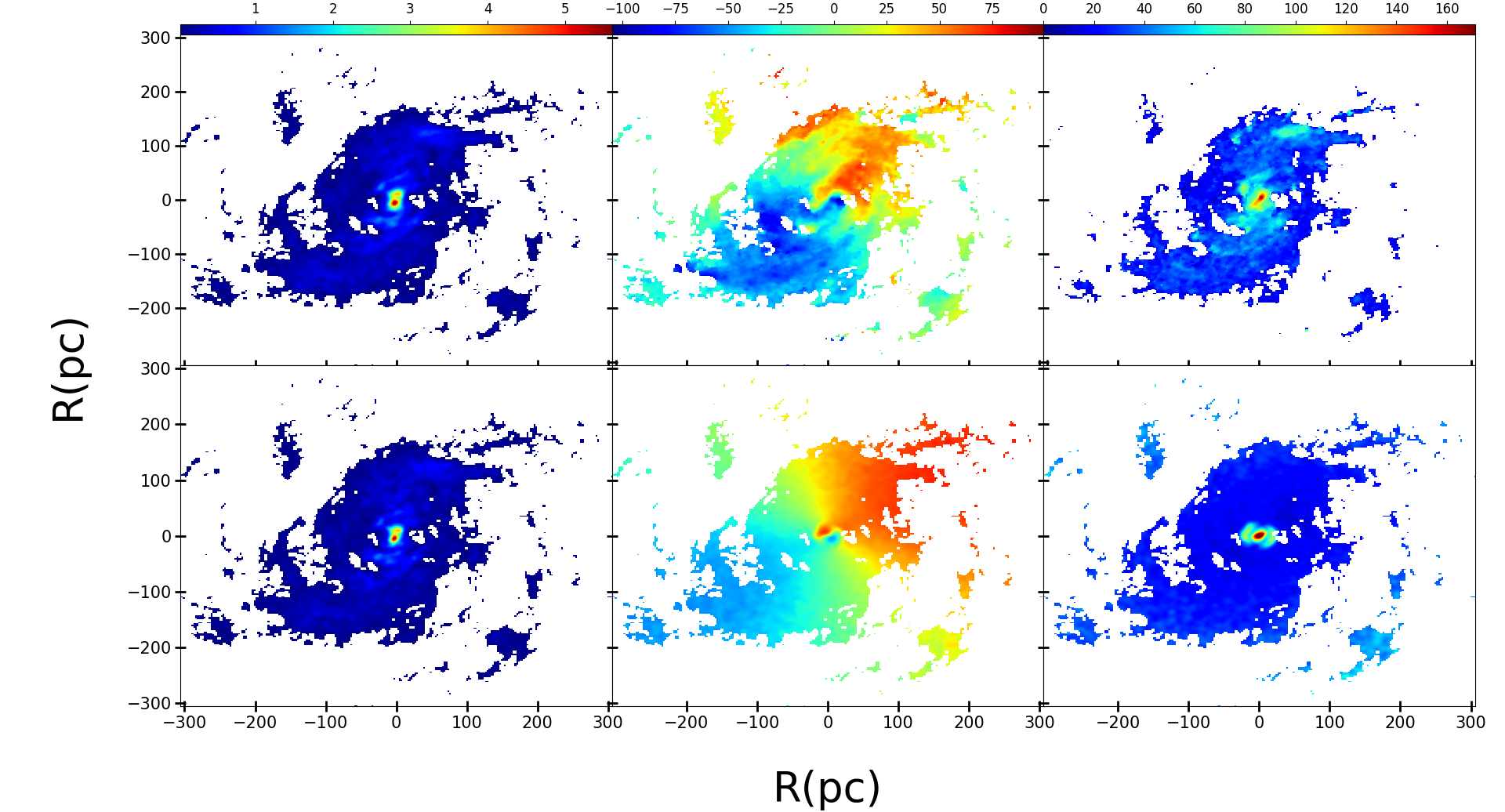}} \hfill%
\subfloat{}{\includegraphics[width=0.49\textwidth]{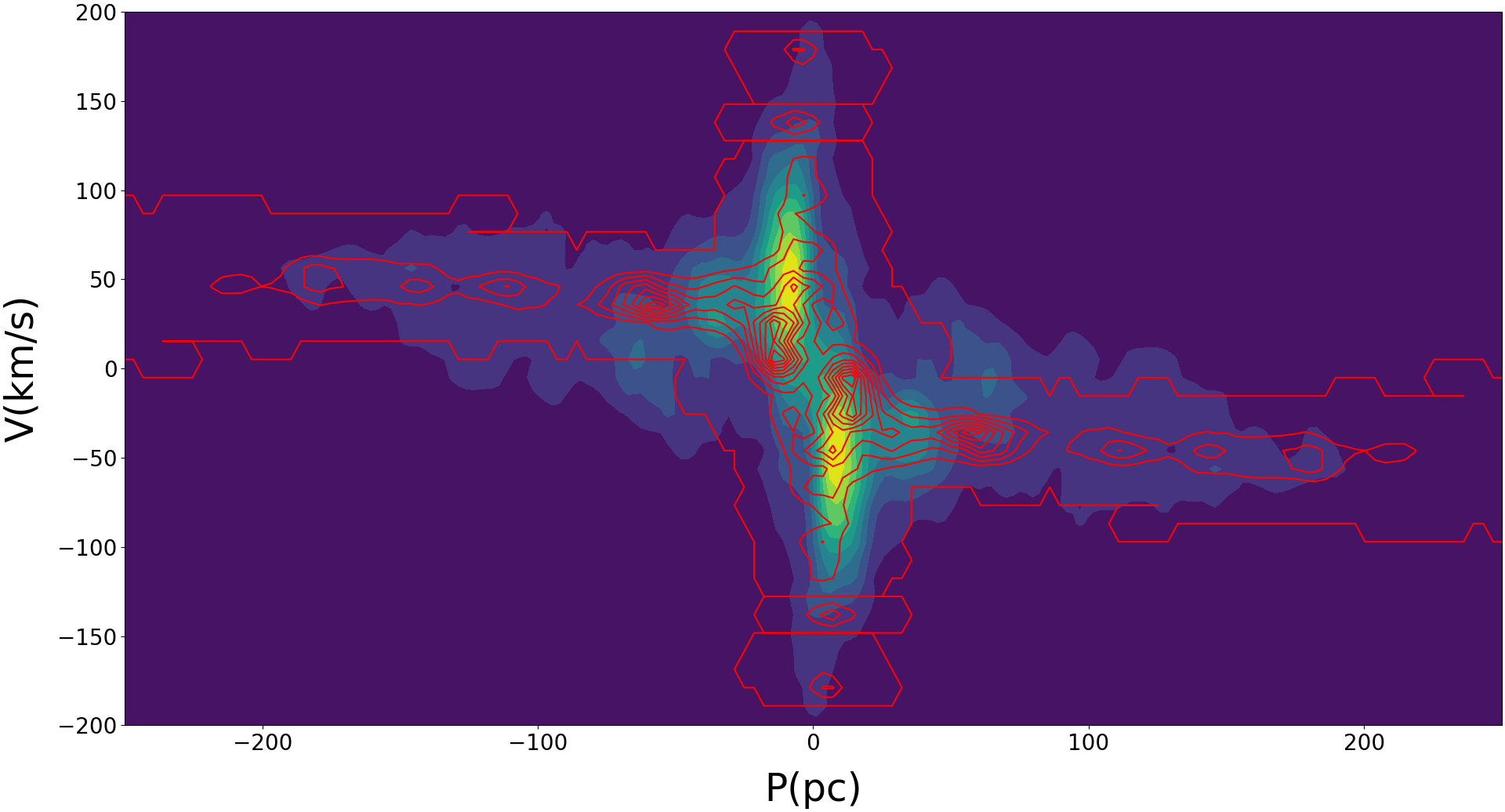}} 
\caption{Same as \ref{fig:NGC4388_mom} for NGC~5643.}
\label{fig:NGC5643_mom}
\end{figure*}

\begin{figure*}[h]
\centering
\subfloat{}{\includegraphics[width=0.49\textwidth]{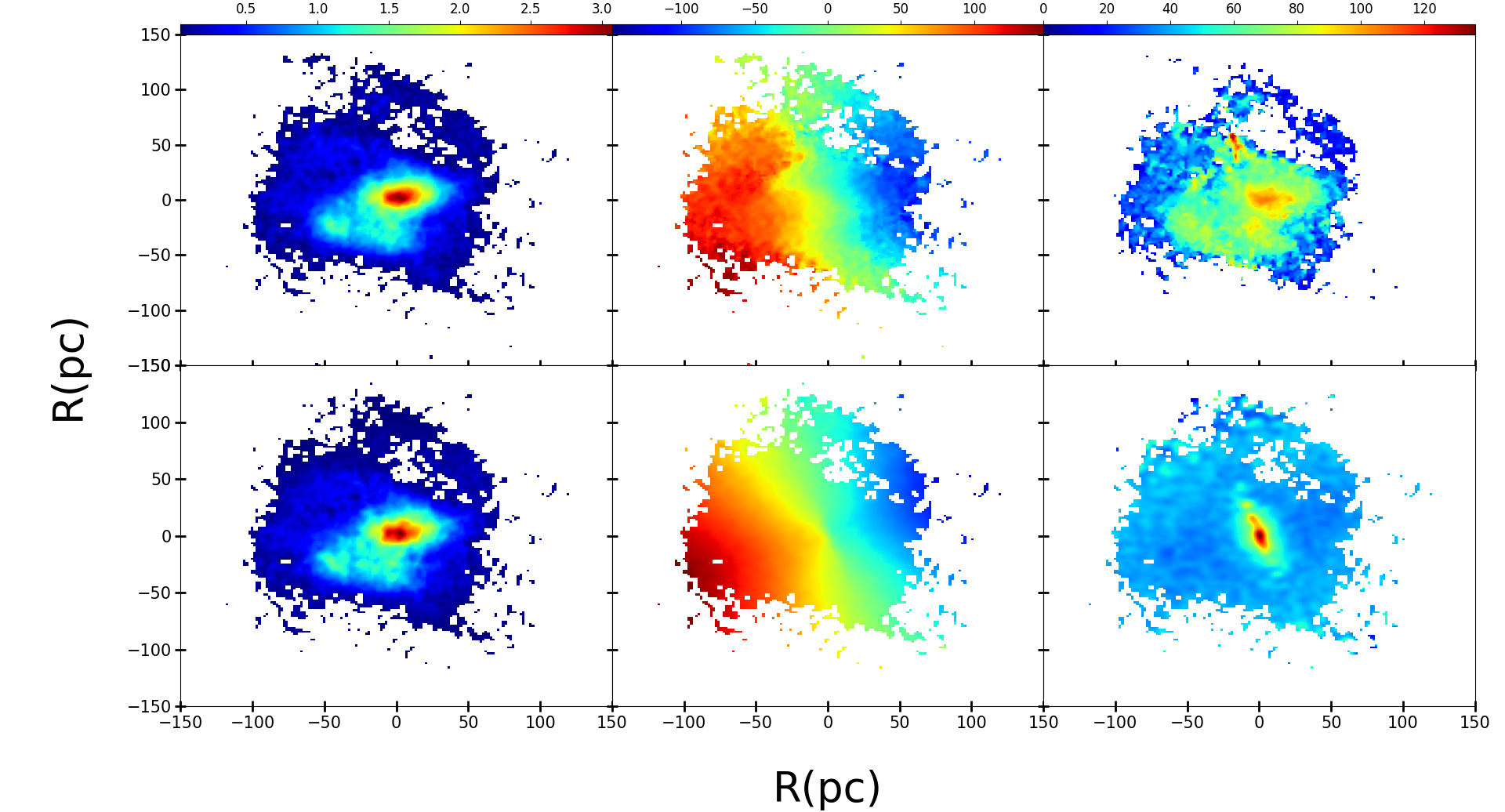}} \hfill%
\subfloat{}{\includegraphics[width=0.49\textwidth]{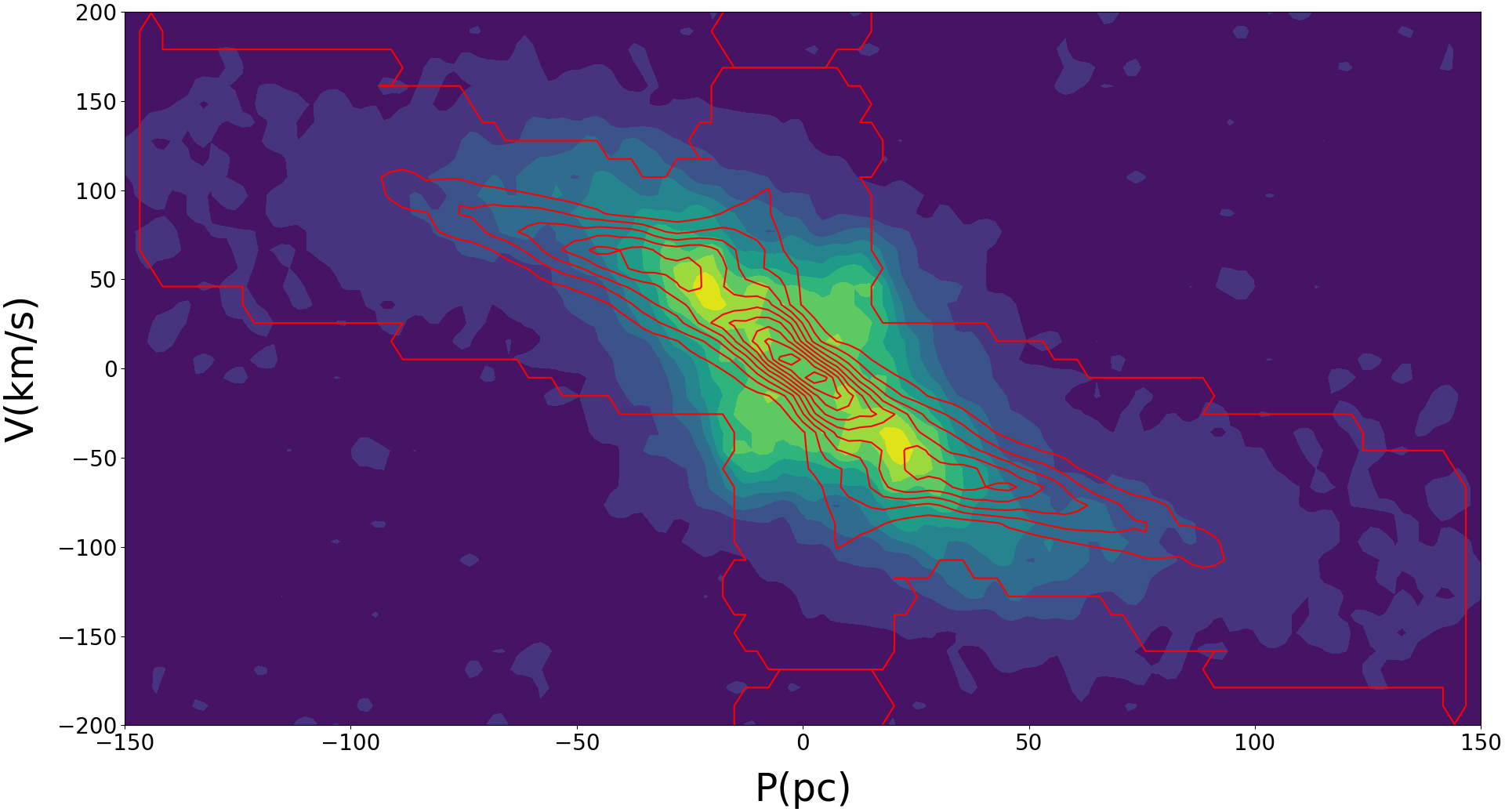}} 
\caption{Same as \ref{fig:NGC4388_mom} for NGC~6300.}
\label{fig:NGC6300_mom}
\end{figure*}

\begin{figure*}[h]
\centering
\subfloat{}{\includegraphics[width=0.49\textwidth]{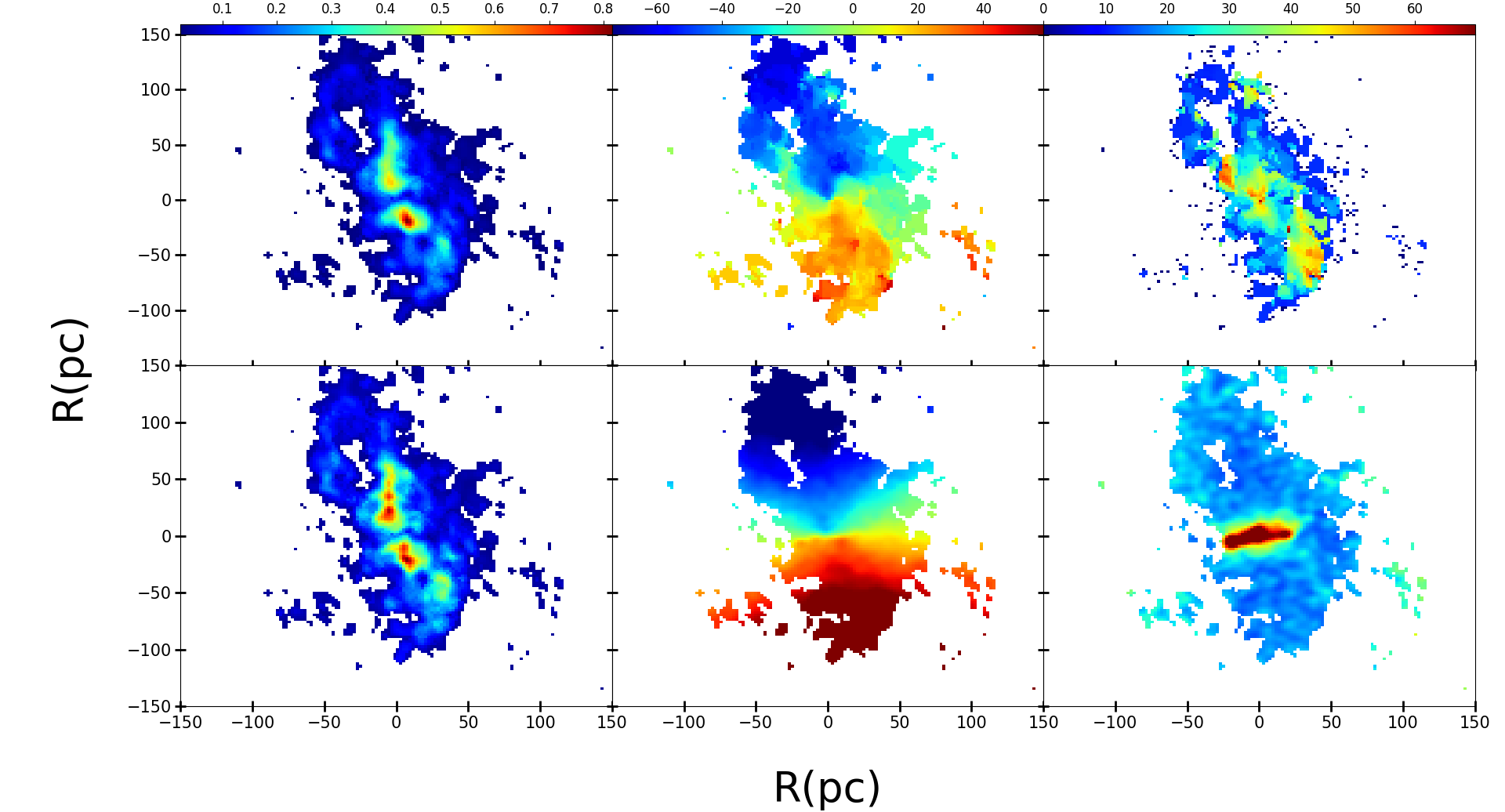}} \hfill%
\subfloat{}{\includegraphics[width=0.49\textwidth]{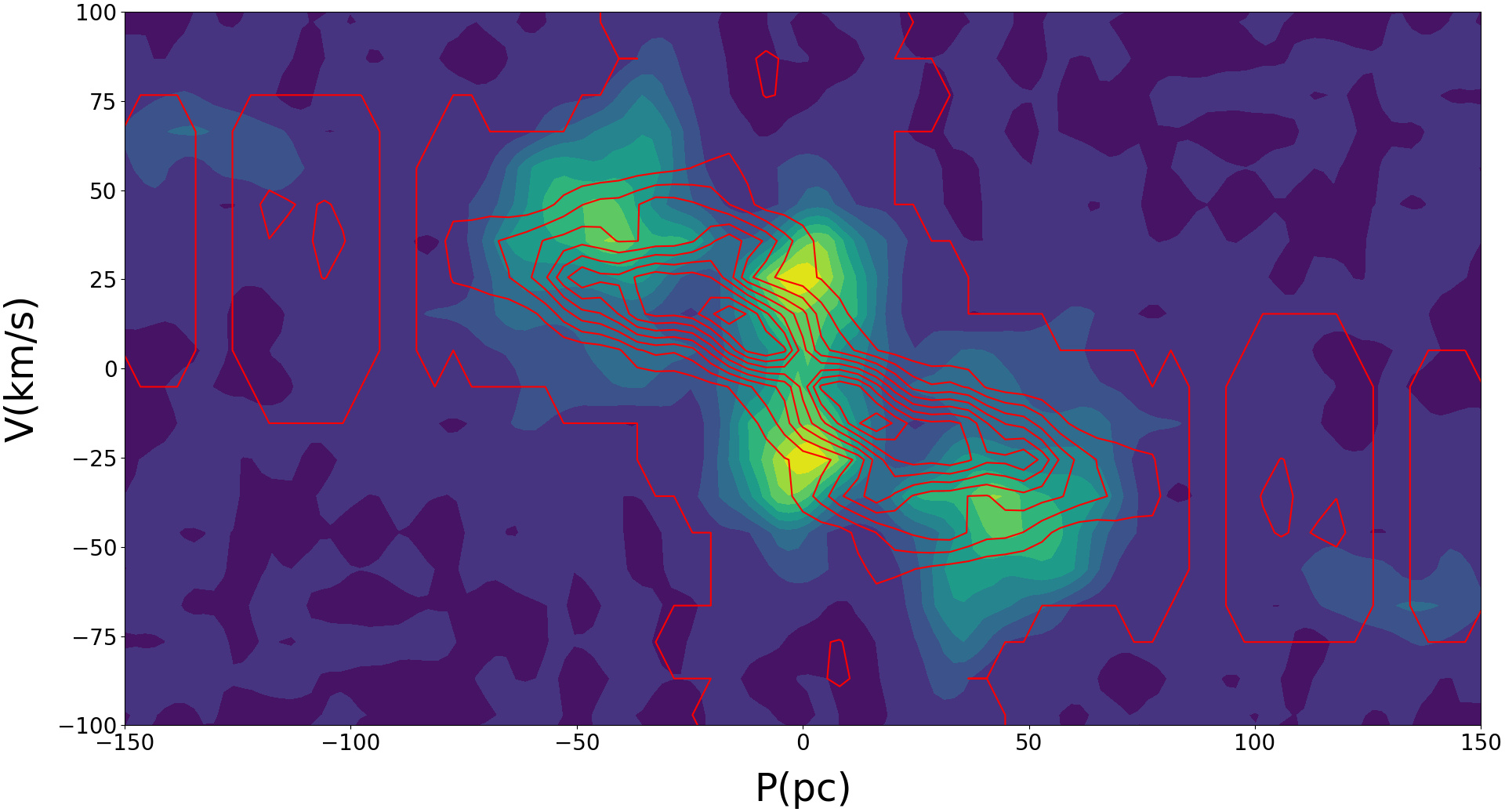}} 
\caption{Same as \ref{fig:NGC4388_mom} for NGC~7314.}
\label{fig:NGC7314_mom}
\end{figure*}

\begin{figure*}[h]
\centering
\subfloat{}{\includegraphics[width=0.49\textwidth]{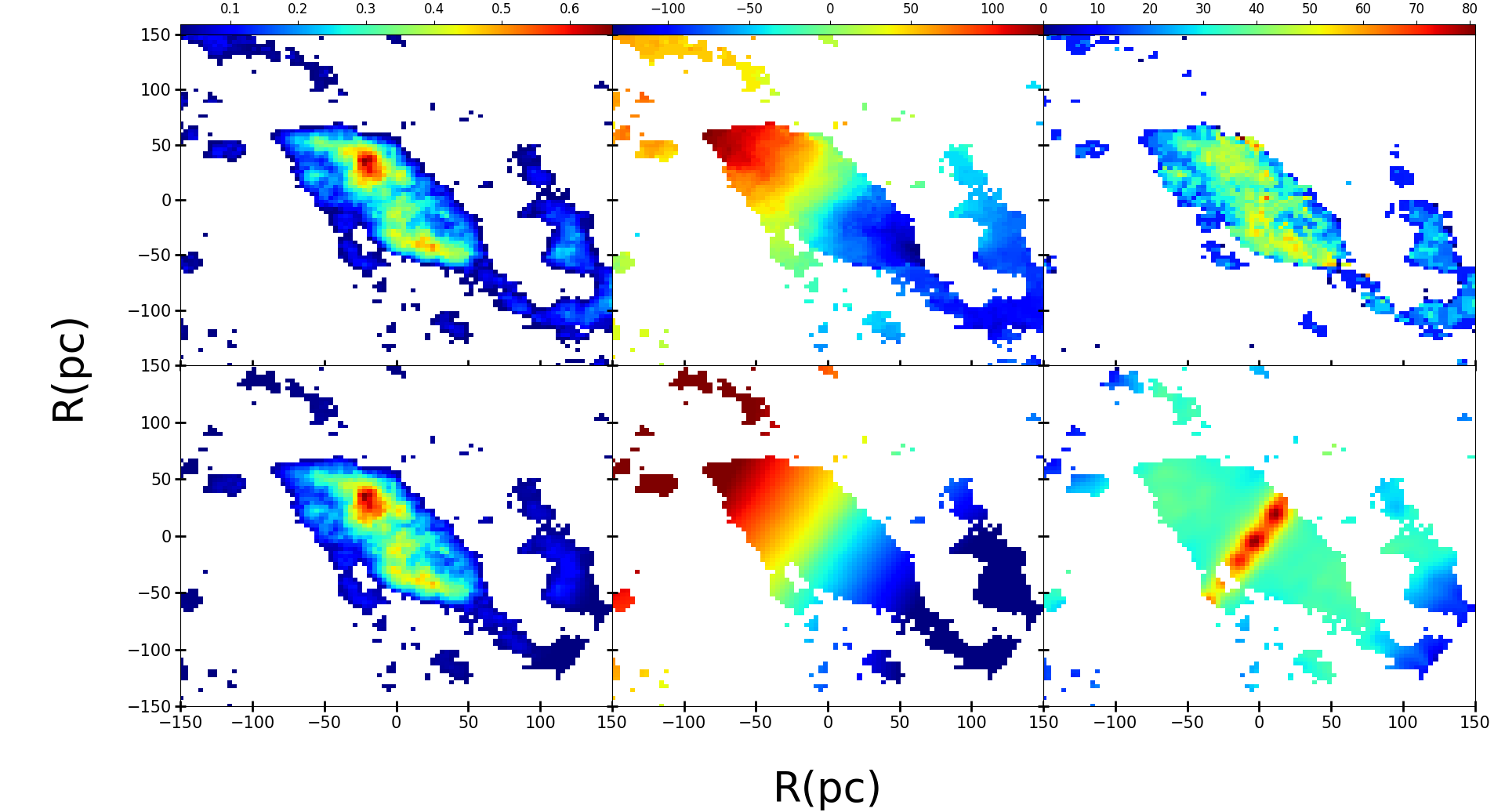}} \hfill%
\subfloat{}{\includegraphics[width=0.49\textwidth]{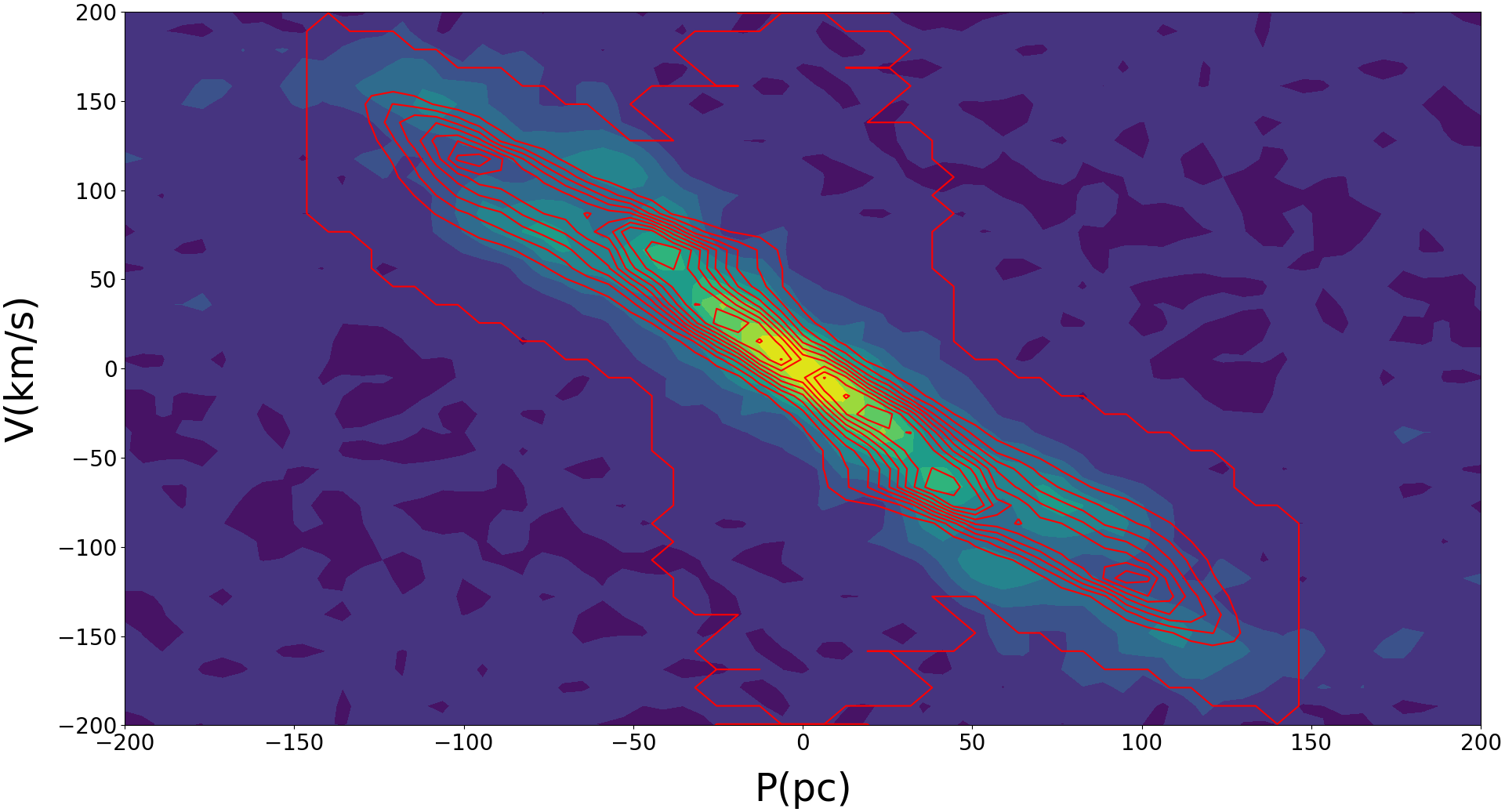}} 
\caption{Same as \ref{fig:NGC4388_mom} for NGC~7465.}
\label{fig:NGC7465_mom}
\end{figure*}

\begin{figure*}[h]
\centering
\subfloat{}{\includegraphics[width=0.49\textwidth]{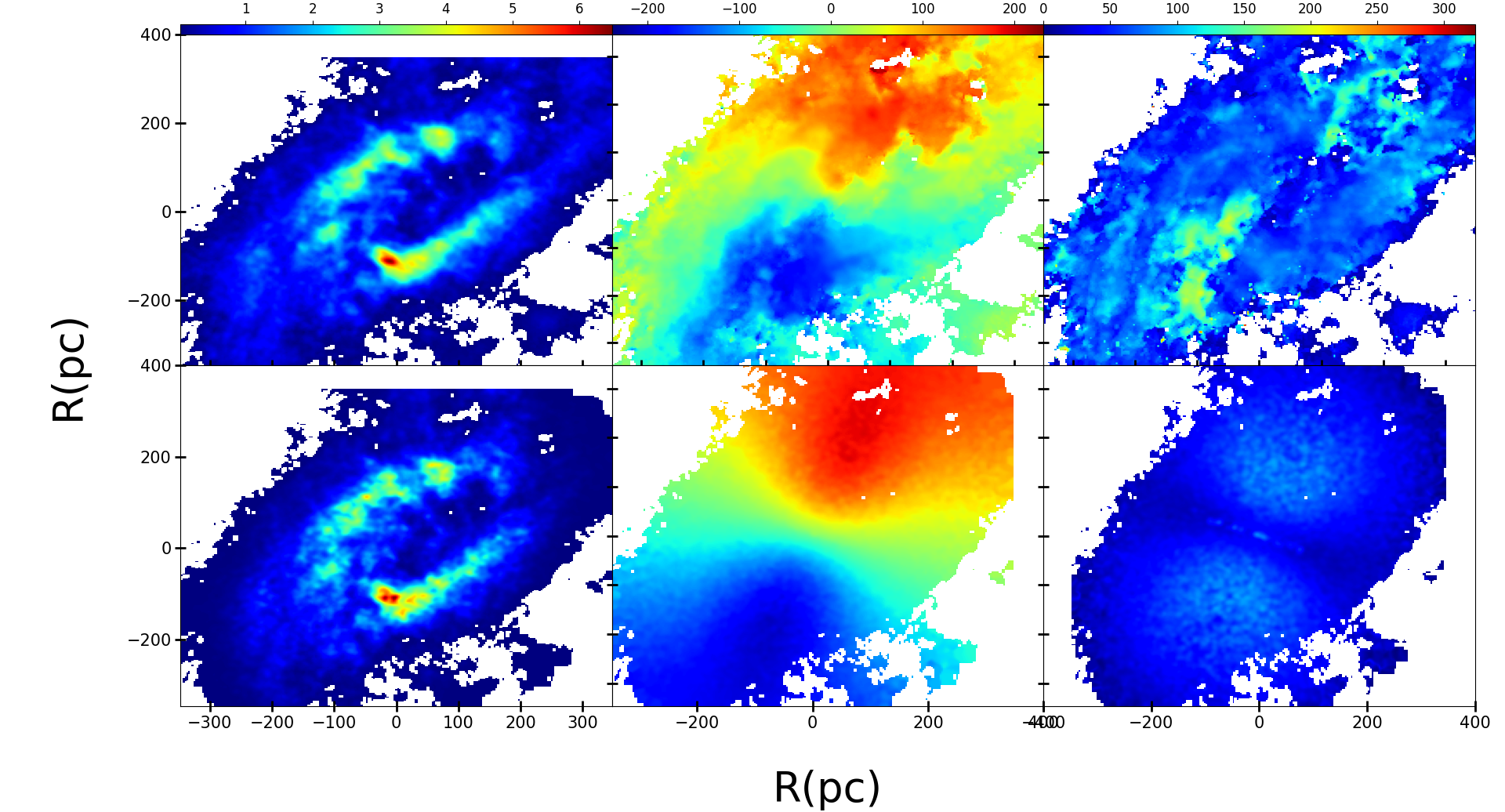}} \hfill%
\subfloat{}{\includegraphics[width=0.49\textwidth]{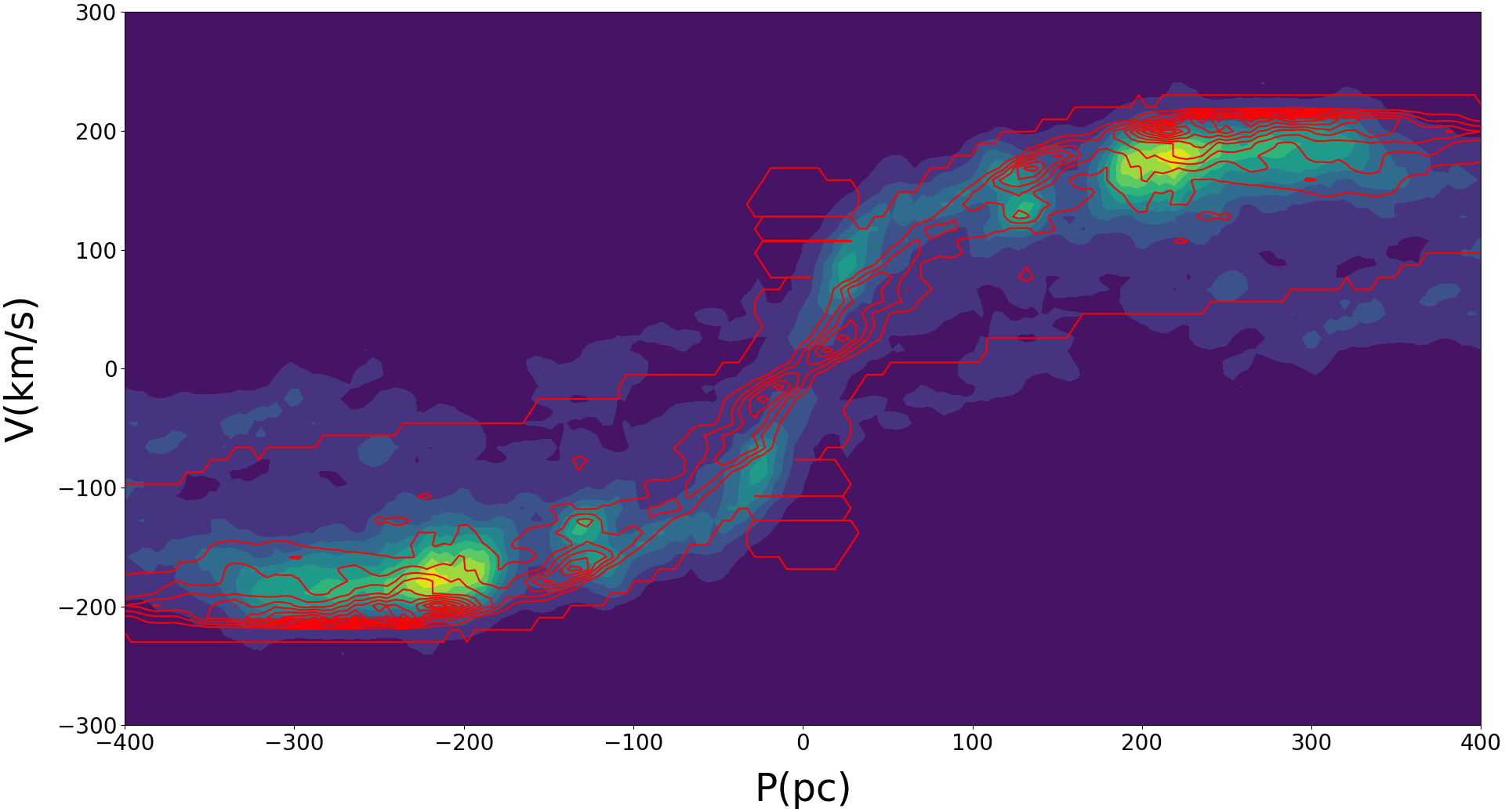}} 
\caption{Same as \ref{fig:NGC4388_mom} for NGC~7582.}
\label{fig:NGC7582_mom}
\end{figure*}

 \FloatBarrier
\begin{table*}[h]
\caption{\label{table:fits_params}Parameters of the moment maps fits}
	\vspace{-0cm}
	\hspace*{1.1cm}
	\centering
	\begin{adjustbox}{width=0.8\textwidth, left}
	\renewcommand{\arraystretch}{1.3}
	\scriptsize
	\begin{tabular}{cccccccc}
		\toprule
		PARAMETER      & NGC4388 & NGC5506 & NGC5643  & NGC6300   & NGC7314 & NGC7465 & NGC7582   \\
		\midrule        
		$M_{\mathrm{bulb}}$           & 0.40    & 0.20    & 0.10    & 0.25     & 0.10    & 0.60    & 0.45       \\
		$R_{\mathrm{bulb}}$           & 0.10    & 0.20    & 0.10    & 0.10     & 0.10    & 0.15    & 0.10       \\
		$H_{\mathrm{bulb}}$           & 0.1     & 0.1     & 0.1     & 0.1      & 0.1     & 0.1     & 0.1       \\
		$M_{\mathrm{gal}}$            & 8.0     & 4.0     & 8.0     & 9.0      & 3.0     & 1.0     & 5.0       \\
		$R_0$                         & 3.0     & 3.0     & 1.0     & 3.0      & 3.0     & 10.0    & 8.0       \\
		$H_t$                         & 0.001   & 0.001   & 0.001   & 0.003    & 0.001   & 0.002   & 0.001     \\
		log $M_{\mathrm{BH}}$         & 6.61    & 6.63    & 6.99    & 6.39     & 6.41    & 6.62    & 6.40      \\
		$RAPM$                        & 0.020   & 0.021   & 0.020   & 0.021    & 0.020   & 0.020   & 0.020     \\
		$R_{\mathrm{gas}}$            & 0.48    & 0.50    & 0.55    & 0.20     & 0.20    & 0.20    & 0.60       \\
		$H_{\mathrm{gas}}$            & 0.12    & 0.012   & 0.011   & 0.03     & 0.02    & 0.12    & 0.08      \\
		$PA$                          & 10      & 0       & [50,150]& 160      & 80      & 30      & 110 \\
		$IA$                          & 79      & 80      & 29      & 57       & 55      & 55      & 59   \\
		\bottomrule
		\end{tabular}
    \end{adjustbox}
    \tablefoot{Table of the parameters used to obtain the fits of the observed moment maps shown in the figures \ref{fig:NGC4388_mom} to \ref{fig:NGC7582_mom}. The 12 parameters are described in the fourth paragraph of the \ref{subsection:numerical_simulation}. For each galaxy, we choose the \mbh~ given by the estimations made with moment maps model.
    In the case of NGC~5643, we have two different $PA$ in order to model the change of $PA$ seen in the observation. The $RAPM$ parameter has no units, 
    $M_{\mathrm{bulb}}$ and $M_{\mathrm{gal}}$ are in units of $2.25\times 10^9 M_{\odot}$, $M_{\mathrm{BH}}$  in $log(M/M_{\odot})$, $PA$ and $IA$ are in degrees and
    $R_{\mathrm{bulb}}$, $H_{\mathrm{bulb}}$, $R_0$, $H_t$, $R_{\mathrm{gas}}$ and $H_{\mathrm{gas}}$ are in kpc.} 
\end{table*}


 \section{RE of the \texorpdfstring{\mbh~}{} predictions as a function of free parameters} \label{app:moderr}

\begin{figure*}[h]
\centering
\hspace*{-1.25cm}
\includegraphics[width=0.7\linewidth,height=10cm,keepaspectratio]{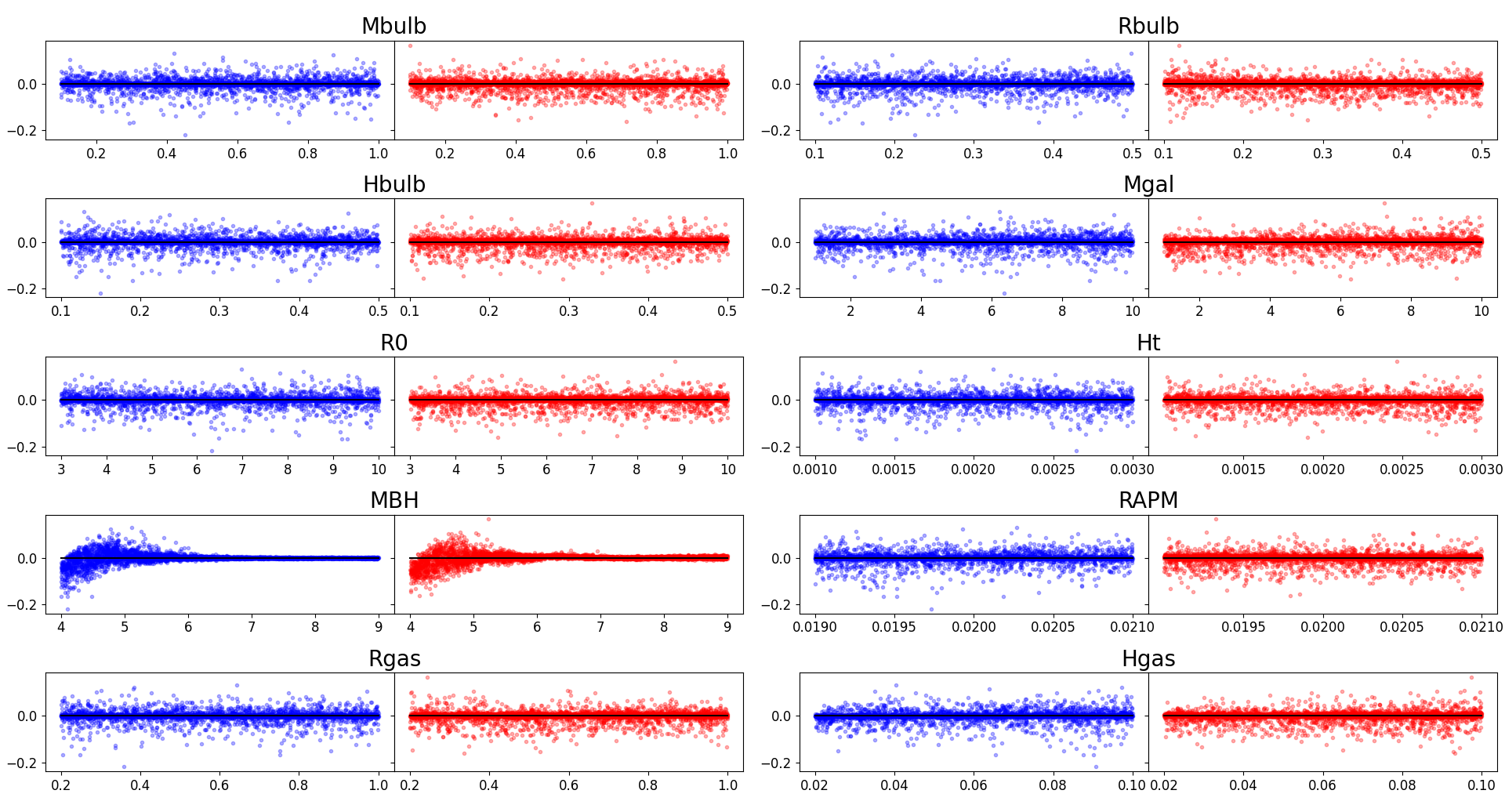}
\caption{NGC~6300's model RE on the \mbh~ estimations made on the test set as a function of the simulation parameter's values. We represented in blue the error of the model using the PVD and the one using the first-moment maps in red. 
The vertical axes represent the RE and the horizontal axes the parameter's values:
$RAPM$ has no units, $M_{\mathrm{bulb}}$ and $M_{\mathrm{gal}}$ are in units of $2.25\times10^9 M_{\odot}$, \mbh~ is in $\mathrm{log}(M/M_{\odot})$ and $R_{\mathrm{bulb}}$, $H_{\mathrm{bulb}}$, R0, Ht, $R_{\mathrm{gas}}$ and $H_{\mathrm{gas}}$ are in kpc.}
\label{fig:err6300_details}
\end{figure*}

 \FloatBarrier
 \section{Models 2D histograms of their RE} \label{app:hist2d}

\begin{figure*}[h]
    \centering
     \includegraphics[width=0.7\linewidth,height=10cm,keepaspectratio]{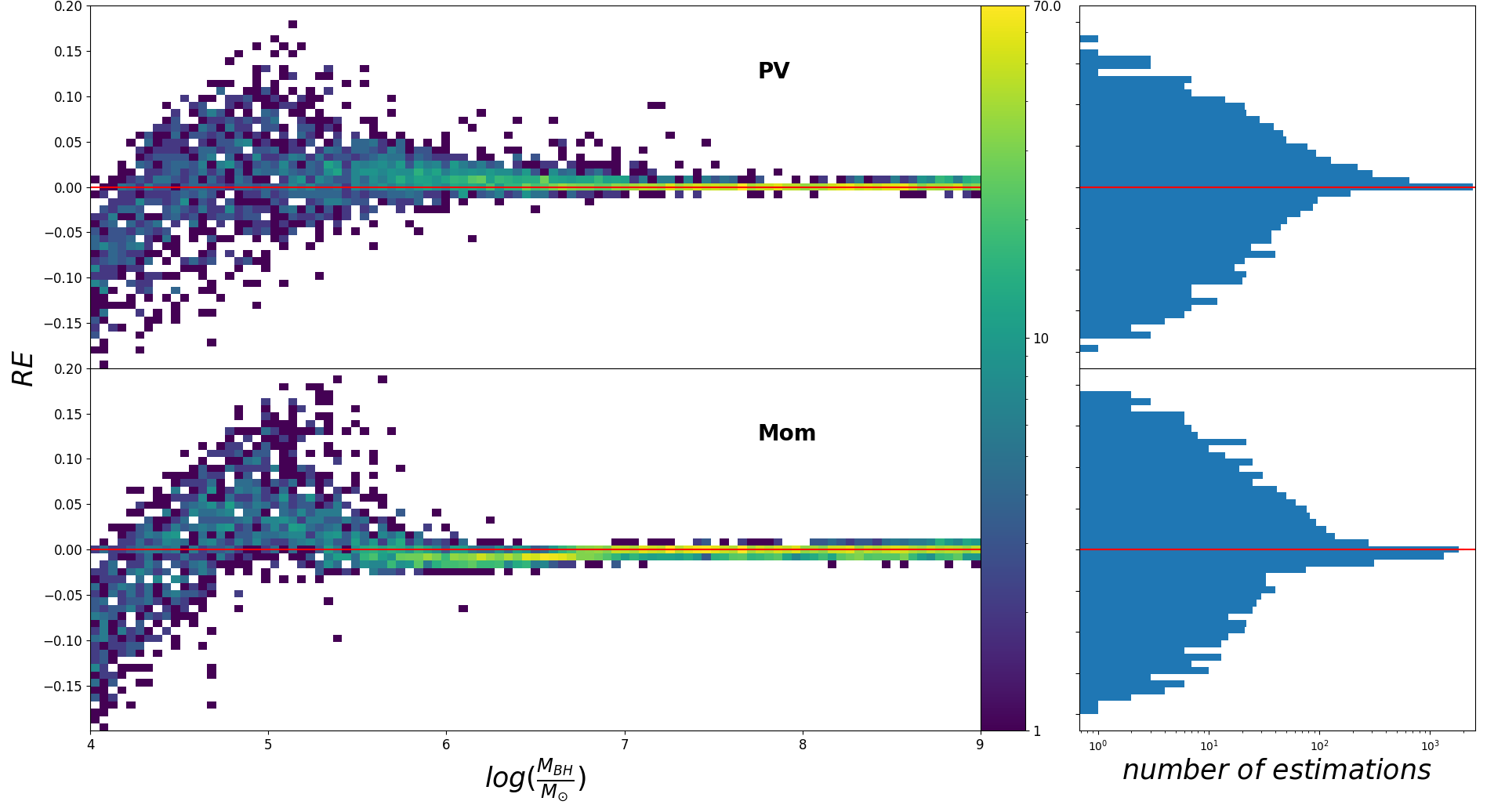}
    \caption{Two-dimensional and one-dimensional histograms of the RE of our models when estimating the \mbh~ of the test set. 
    On the left, we plotted the RE of NGC~4388 models when estimating the \mbh~ of the test set as 2D histograms. On the right, we plotted the same RE but as a 1D histogram. 
    On top, we plotted the PVD model RE and at the bottom, the first-moment map model RE.
    The y-axis represents the RE values. The x-axis of the 2D histograms is in $log(\frac{M_{\mathrm{BH}}}{M_{\odot}})$ and the 1D histograms axis is in number of estimations.}
    \label{fig:4388_hist2D}
\end{figure*}

\begin{figure*}[h]
    \centering
     \includegraphics[width=0.7\linewidth,height=10cm,keepaspectratio]{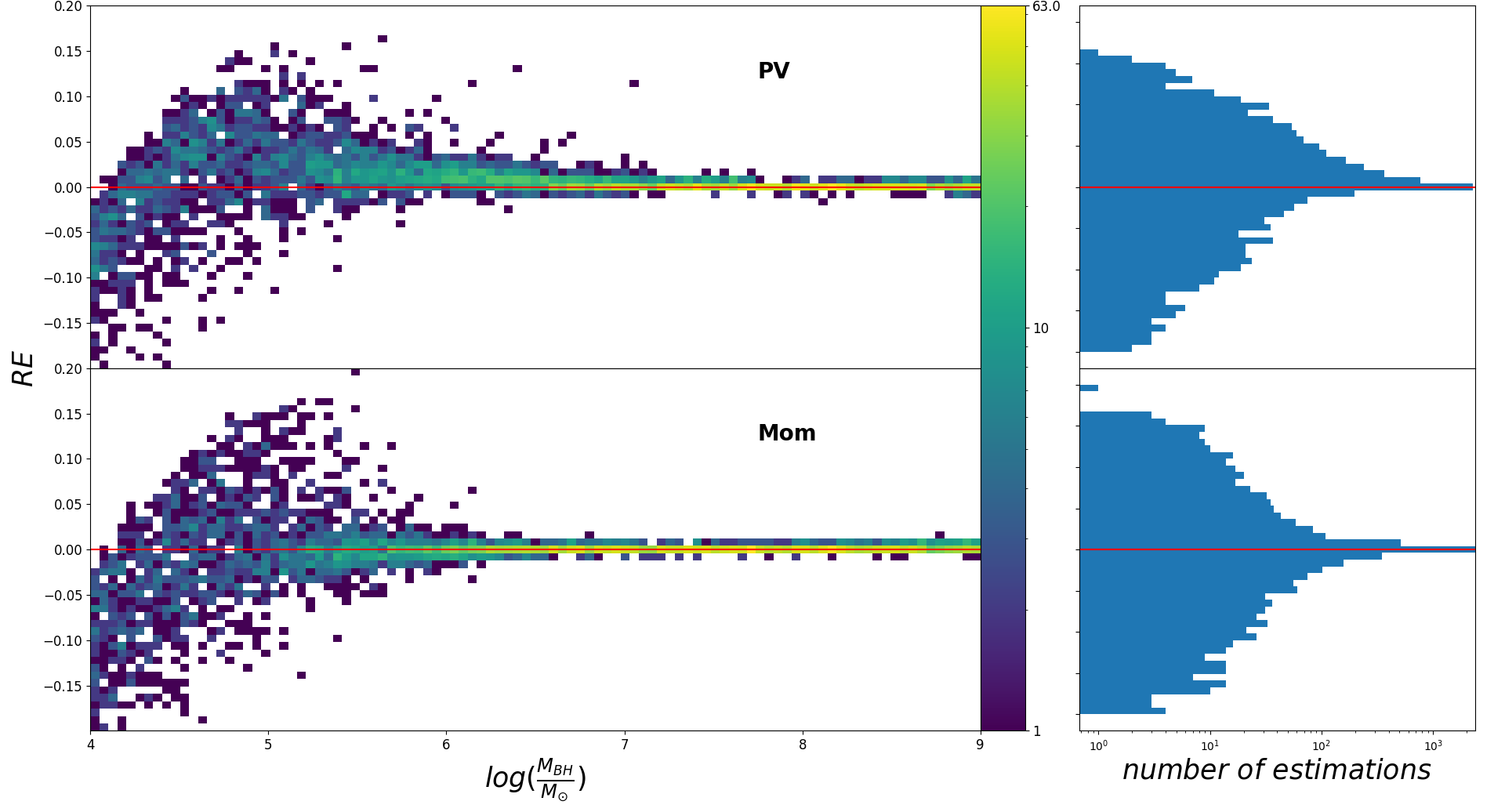}
    \caption{Same as \ref{fig:4388_hist2D} for NGC~5506.}
    \label{fig:5506_hist2D}
\end{figure*}

\begin{figure*}[h]
    \centering
     \includegraphics[width=0.7\linewidth,height=10cm,keepaspectratio]{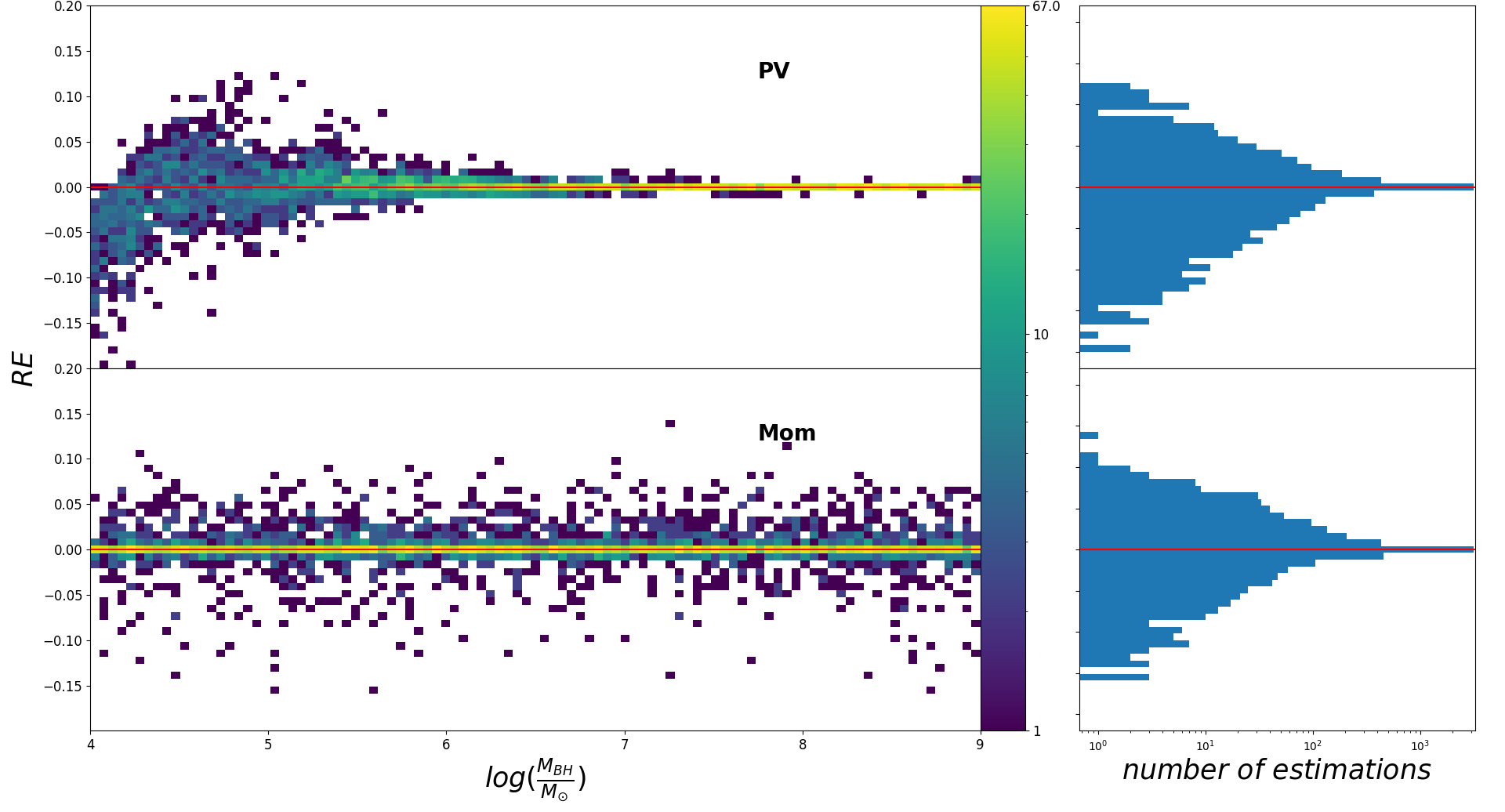}
    \caption{Same as \ref{fig:4388_hist2D} for NGC~5643.}
    \label{fig:5643_hist2D}
\end{figure*}

\begin{figure*}[h]
    \centering
     \includegraphics[width=0.7\linewidth,height=10cm,keepaspectratio]{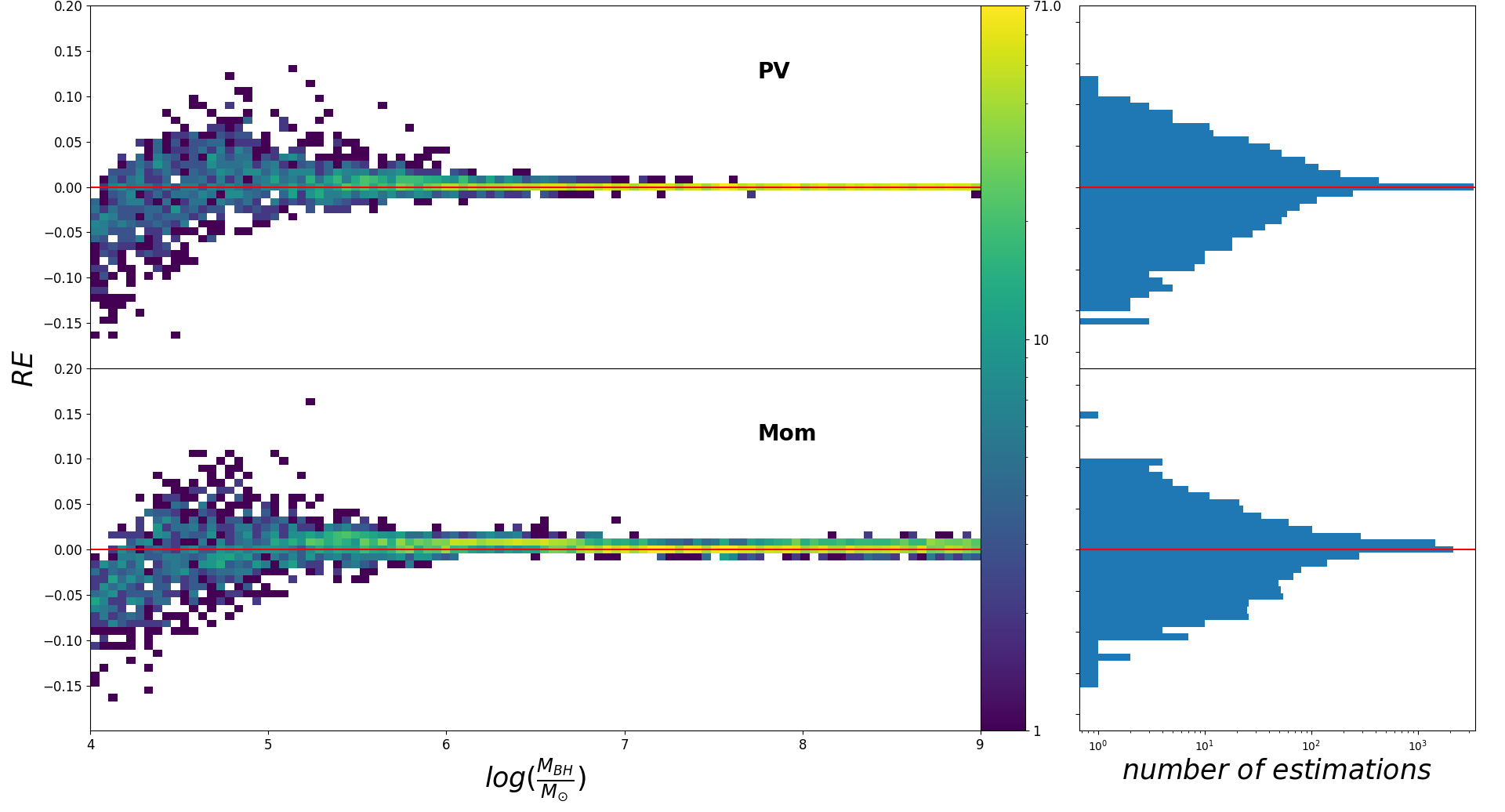}
    \caption{Same as \ref{fig:4388_hist2D} for NGC~6300.}
    \label{fig:6300_hist2D}
\end{figure*}

\begin{figure*}[h]
    \centering
     \includegraphics[width=0.7\linewidth,height=10cm,keepaspectratio]{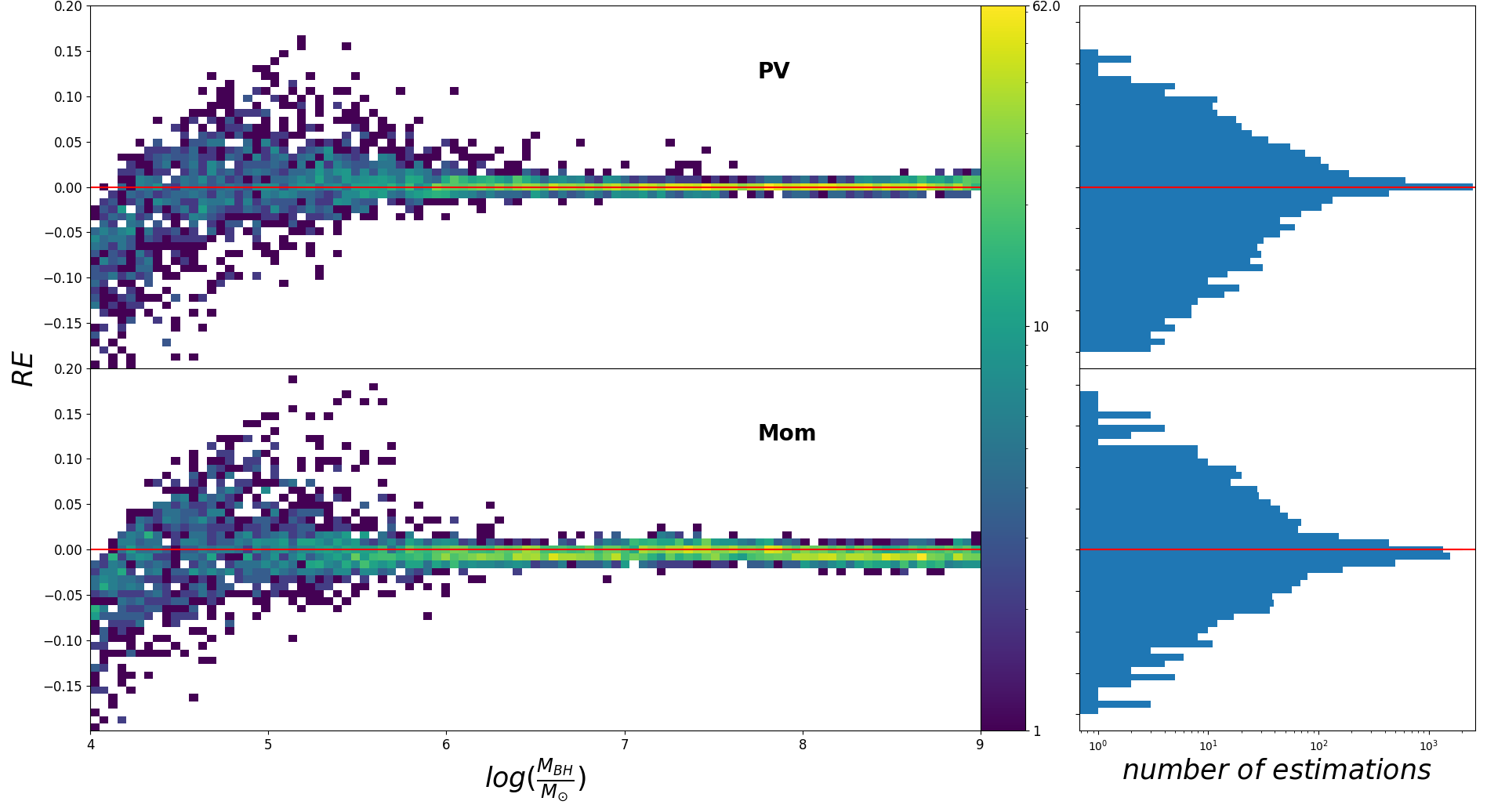}
    \caption{Same as \ref{fig:4388_hist2D} for NGC~7314.}
    \label{fig:7314_hist2D}
\end{figure*}

\begin{figure*}[h]
    \centering
     \includegraphics[width=0.7\linewidth,height=10cm,keepaspectratio]{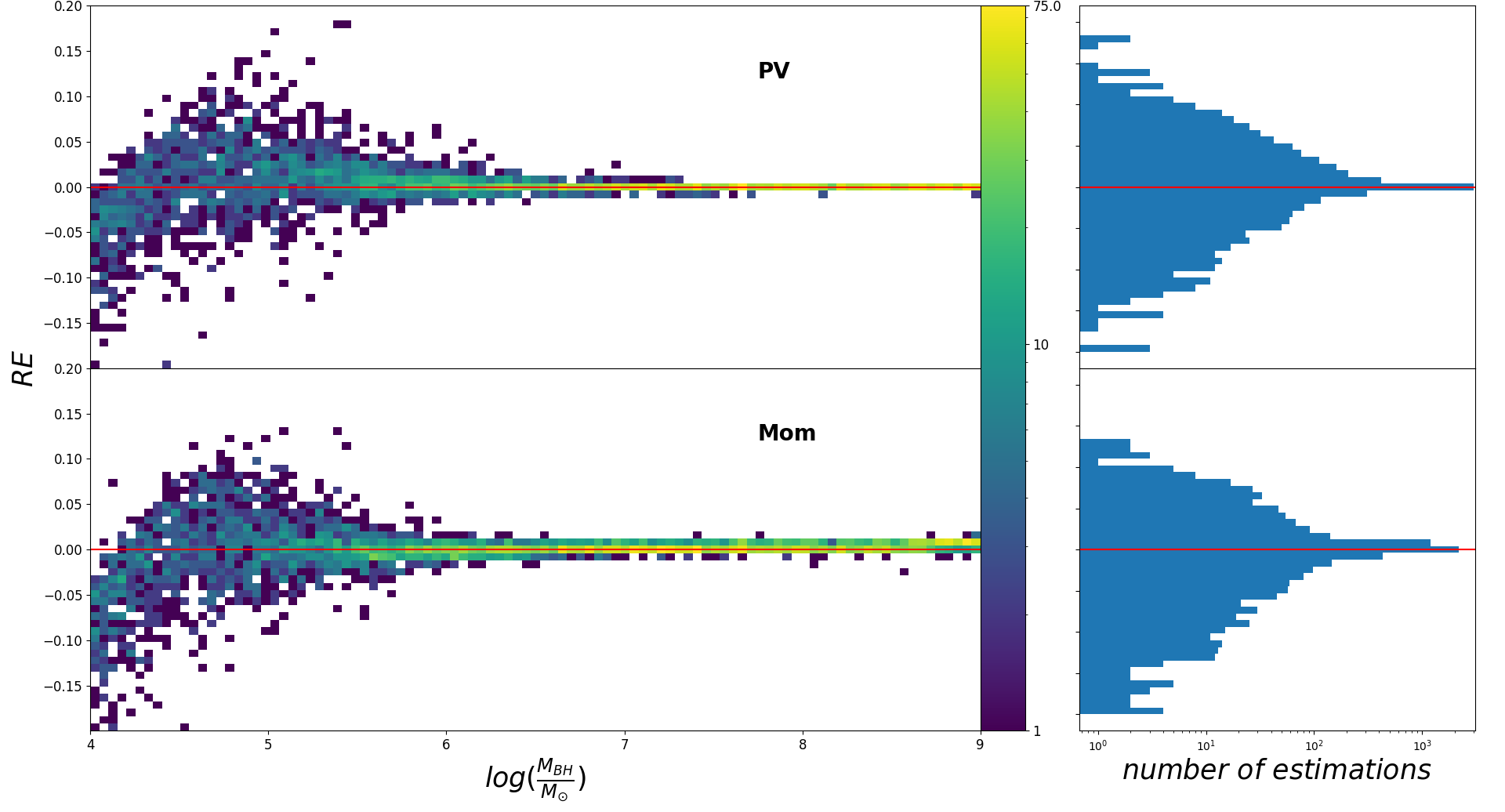}
    \caption{Same as \ref{fig:4388_hist2D} for NGC~7465.}
    \label{fig:7465_hist2D}
\end{figure*}

\begin{figure*}[h]
    \centering
     \includegraphics[width=0.7\linewidth,height=10cm,keepaspectratio]{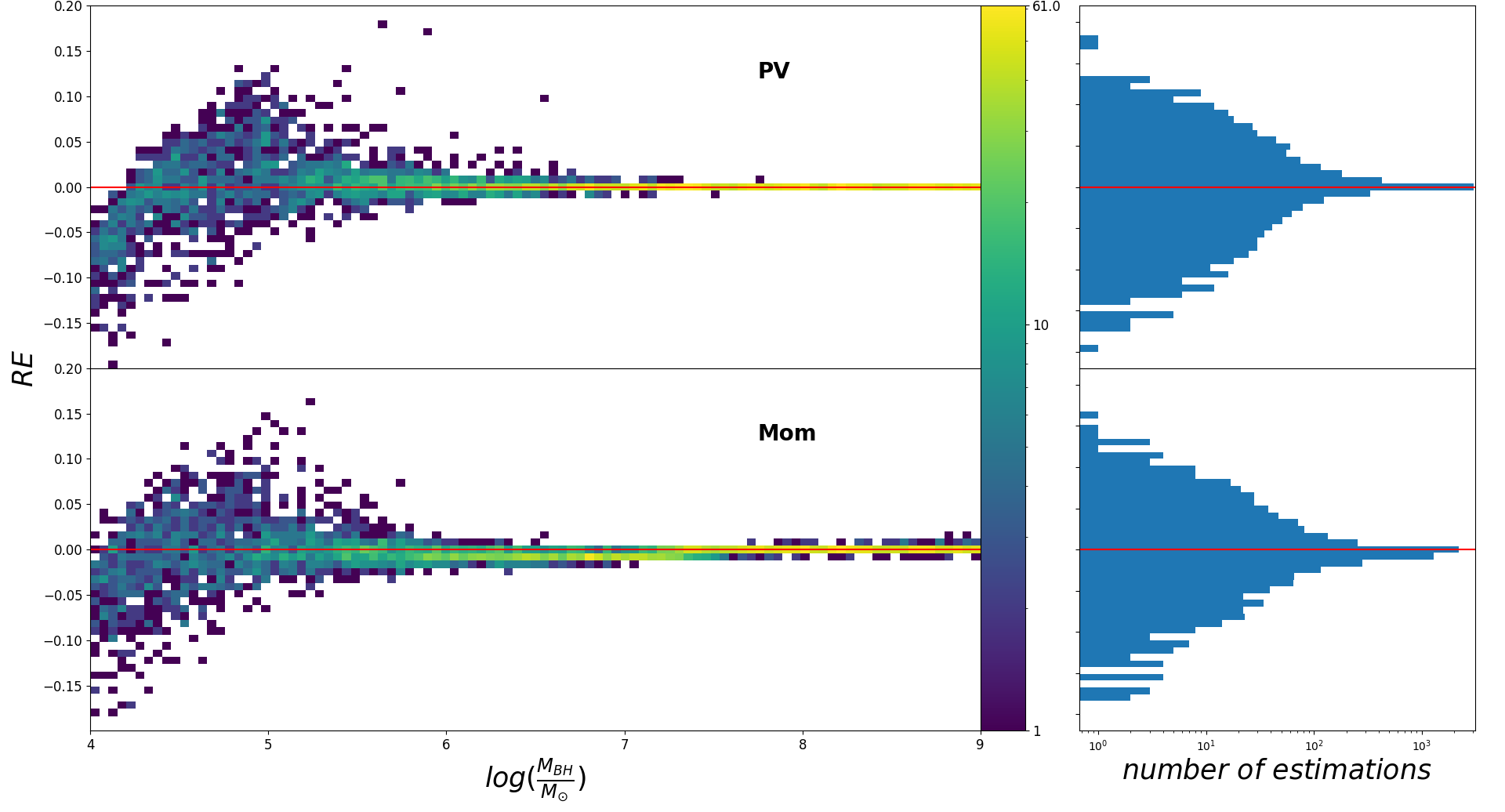}
    \caption{Same as \ref{fig:4388_hist2D} for NGC~7582.}
    \label{fig:7582_hist2D}
\end{figure*}

 \end{appendix}

\end{document}